\documentclass[final,leqno]{siamltex704}

\usepackage{amsmath}
\usepackage{graphicx}
\usepackage{amssymb}
\usepackage{longtable}
\usepackage[nocompress,space]{cite}
\usepackage{exscale}

\newcommand{\A}{{\mathcal A}}
\newcommand{\B}{{\mathcal B}}
\newcommand{\C}{{\mathcal C}}
\newcommand{\D}{{\mathcal D}}

\newcommand{\G}{{\mathcal G}}

\newcommand{\vecu}{{\sf u}}
\newcommand{\vecv}{{\sf v}}

\newcommand{\bbZ}{{\mathbb{Z}}}

\renewcommand{\ldots}{\ensuremath{\dotsc}}

\newcommand{\nn}{\nonumber}

\newcommand{\bd}{\begin{document}}
\newcommand{\ed}{\end{document}}
\newcommand{\clearabove}{\rule{0pt}{9pt}}
\newcommand{\clearbelow}{\rule[-5pt]{0pt}{9pt}}
\newcommand{\clearboth}{\rule[-5pt]{0pt}{14pt}}
\newcommand{\spaceouttable}{\renewcommand{\arraystretch}{1.2}}
\renewcommand{\topfraction}{1.0} 
\renewcommand{\bottomfraction}{1.0} 
\renewcommand{\floatpagefraction}{.75}
\renewcommand{\textfraction}{.095} 
\title{Spectrum of Sizes for Perfect Deletion-Correcting~Codes\thanks{Received 
by the editors March 2, 2009;
accepted for publication (in revised form) November 20, 2009;
published electronically February 5, 2010.
\URL sidma/24-1/75131.html}}

\author{Yeow Meng Chee\thanks{Division of Mathematical Sciences,
School of Physical and Mathematical Sciences, Nanyang Technological University,
21~Nanyang Link, Singapore 637371 (ymchee@ntu.edu.sg). This author's research
was supported in part by the National Research Foundation of Singapore
 under research grant NRF-CRP2-2007-03
 and by the Nanyang Technological University under research grant M58110040.}
\and 
Gennian Ge\thanks{Corresponding author. Department of Mathematics,
Zhejiang University, Hangzhou 310027, Zhejiang, People's Republic of China 
(gnge@zju.edu.cn). The author's research 
was supported by the National Outstanding
Youth Science Foundation of China under grant 10825103, the
National Natural Science Foundation of China under grant
10771193, the Specialized Research Fund for the Doctoral Program
of Higher Education, Program for New Century Excellent Talents in
University, and by the Zhejiang Provincial Natural Science Foundation of
China under grant D7080064.}
\and 
Alan C. H. Ling\thanks{Department of Computer Science,
University of Vermont, Burlington, VT 05405
(aling@\allowbreak emba.\allowbreak uvm.\allowbreak edu).
This work was done while the author was on
 sabbatical leave at the Division of Mathematical Sciences, School of Physical and
 Mathematical Sciences, Nanyang Technological University, Singapore.}}

\begin{document}
\slugger{sidma}{2010}{24}{1}{33--55}
\maketitle

\setcounter{page}{33}

\begin{abstract}
One peculiarity with deletion-correcting codes is that perfect
$t$-deletion-correcting codes of the same length over the same
alphabet can have different numbers of codewords, because the balls
of radius $t$ with respect to the Levenshte\u{\i}n distance may be of
different sizes. There is interest, therefore, in determining all
possible sizes of a perfect $t$-deletion-correcting code, given
the
 length $n$ and the alphabet size~$q$. In this
paper, we determine completely the spectrum of possible sizes for
perfect $q$-ary 1-deletion-correcting codes of length three for
all $q$, and perfect $q$-ary 2-deletion-correcting codes of length
four for almost all $q$, leaving only a small finite number of
cases in doubt.
\end{abstract}



\begin{DOI}
10.1137/090751311
\end{DOI}

\pagestyle{myheadings} 
\thispagestyle{plain} 
\markboth
{YEOW MENG CHEE, GENNIAN GE, AND ALAN C.\ H.\ LING}
{PERFECT DELETION-CORRECTING CODES}

\section{Introduction}
\label{sec1}

When communication takes place over an insertion-deletion channel,
data is lost and gained between the source and the receiver at
unknown locations in the data stream. Insertion-deletion channels
are rather commonplace as follows:

\begin{itemize}
\item{\em Magnetic and optical recording\/} \cite{Bours:1994,RothSiegel:1994}:
Random fluctuations in the motion of the recording medium,
during both the read and write processes, cause timing uncertainty in the read-back signal. The
resulting received data stream can miss data at certain locations.

\item{\em Packet-switched communication\/} \cite{SchulmanZuckerman:1999}:
Effects of congestion control protocols and randomness
of packet arrivals give rise to random packet loss.

\item{\em DNA replication\/}
\cite{Marronetal:2003,PrattCornely:2004}:
DNA can undergo deletion and insertion mutations, where
bases are deleted or added, because of
strand slippage and homologous recombination.

\item{\em Music performance\/} \cite{Heijinketal:2000,PalmervandeSande:1993}:
In music performance, pitch errors can occur. For instance, notes can be
played that are not in the score (insertion errors) and notes that are specified in the score are omitted
(deletion errors).
\end{itemize}

\noindent
It is therefore
not surprising that the study of codes for combating deletions has continued to this day
\cite{DaveyMacKay:2001,Fuji-Haraetal:2001,Yin:2001,Shalabyetal:2002,Sloane:2002,SwartFerreira:2003,Klein:2004,Ratzer:2005,Wangetal:2005,Tenengolts:1984,WangYin:2006,DolecekAnantharam:2007,McAvenSafavi-Naini:2007,TonienSafavi-Naini:2007}
since its systematic treatment by Levenshte{\u\i}n
\cite{Levenshtein:1965a,Levenshtein:1965b}
and others
\cite{Sellers:1962,Ullman:1966,Ullman:1967,CalabiHartnett:1969a,CalabiHartnett:1969b,VarshamovTenengolts:1965} in the 1960s.

Let $X$ be a finite set of $q$ elements and $n$ be a positive
integer. A set $\C\subseteq X^n$ is called a {\em$q$-ary code of
length $n$}. The set $X$ is called the {\em alphabet\/} of the code
$\C$ and the elements of $\C$ are called {\em codewords}. The {\em size\/} of $\C$ is $|\C|$, the number of codewords it contains. For
$x\in X^n$ and $0\leq t\leq n$, let $D_t(x)$ denote the set of
{\em$t$-th order descendants}, that is, the set of $y\in X^{n-t}$
that are obtained if any $t$ components are deleted from $x$. The
$q$-ary code $\C\subseteq X^n$ is said to be {\em$t$-deletion-correcting\/} if $D_t(x)\cap D_t(y)=\varnothing$ for
all distinct $x,y\in \C$. We call such a code an {\em$(n,t)_q$-deletion-correcting code\/} and denote it by
$(n,t)_q\text{-DCC}$. An alternative characterization of
deletion-correcting codes is via a metric known as the
Levenshte{\u\i}n distance. The {\em Levenshte{\u\i}n distance\/} between two vectors $x,y\in X^n$, denoted $d_L(x,y)$, is defined
as the smallest number of deletions and insertions needed to
change $x$ to $y$. A code $\C$ is $t$-deletion-correcting if and
only if $d_L(x,y)\geq 2t+1$ for all distinct $x,y\in\C$.
Levenshte{\u\i}n \cite{Levenshtein:1965a} showed that a code
capable of correcting $t$ deletions is also capable of correcting
any combination of up to $t$ deletions and insertions. For this
reason, deletion-correcting codes are also often called
insertion/deletion-correcting codes.

Motivated by the definition of a perfect error-correcting code,
Levenshte{\u\i}n \cite{Levenshtein:1991} defines an
$(n,t)_q\text{-DCC}$ $\C\subseteq X^n$ to be {\em perfect\/} if the
balls $D_t(x)$, $x\in\C$, partition $X^{n-t}$. However, as
observed by Sloane \cite{Sloane:2002}, unlike in the case of
error-correcting codes, perfect $t$-deletion-correcting codes of
the same length over the same alphabet can have different numbers
of codewords---an initially surprising fact that is explained by
the balls $D_t(x)$ having different sizes for different codewords
$x$.
It is interesting, therefore, to ask for the possible sizes that a perfect
$(n,t)_q\text{-DCC}$ can have.

\section{State of affairs}
\label{sec2}

Define
\begin{align*}
A_q(n,t) = \max\{|\C|: & \text{ $\C$ is an
$(n,t)_q\text{-DCC}$}\}.
\end{align*}
If $\C$ is an $(n,t)_q\text{-DCC}$ such that $|\C|=A_q(n,t)$, we
say that $\C$ is {\em optimal}.

It is easy to see that for all $q\geq 2$ and $n\geq 1$, we have
\begin{equation*}
A_q(n,t)=\begin{cases}
1&\text{if $t=n$}, \\
q&\text{if $t=n-1$}.
\end{cases}
\end{equation*}

If we define the {\em spectrum of sizes of a perfect\/} $(n,t)_q\text{-DCC}$ as
\begin{equation*}
{\rm Spec}(q,n,t) = \{|\C|:\text{$\C$ is a perfect $(n,t)_q\text{-DCC}$}\},
\end{equation*}
then it is also easy to show that
\begin{equation*}
{\rm Spec}(q,n,t)=\begin{cases}
\{1\}&\text{if $t=n$}, \\
\{\lceil q/n\rceil,\ldots,q\}&\text{if $t=n-1$.}
\end{cases}
\end{equation*}

Hence, the problem of determining ${\rm Spec}(q,n,t)$ for $t\in\{n-1,n\}$ is well solved.
We consider the case $t=n-2$ in this paper. This case is nontrivial and has been treated
by various researchers
\cite{Levenshtein:1991,Bours:1994,Bours:1995,Mahmoodi:1998,Yin:2001,Shalabyetal:2002,Klein:2004,WangYin:2006}.
The following result on the size of an
optimal $(n,n-2)_q\text{-DCC}$ was obtained by Bours~\cite{Bours:1995}.

\begin{theorem}[Bours \cite{Bours:1995}]
\label{bound}
Define
\begin{equation*}
DU(q,n) := \left\lfloor \frac{q}{n} \left\lfloor
\frac{2(q-1)}{n-1}
\right\rfloor \right\rfloor+q.
\end{equation*}
Then $A_q(n,n-2)\leq DU(q,n)$ for all $q\geq 2$ and $n\geq 2$.
\end{theorem}

The existence of perfect $(3,1)_q\text{-DCCs}$ has been settled by
Levenshte{\u\i}n \cite{Levenshtein:1991}, where all the codes are
 also optimal. Bours \cite{Bours:1995} also proved the existence of
two more classes of perfect $(n,n-2)_q\text{-DCCs}$, but their
codes are not always optimal.

\begin{theorem}[Bours \cite{Bours:1995}]
\label{Bours-1} 
The following codes exist.

\begin{remunerate}
\item A perfect $(4,2)_q\text{-DCC}$ of size
$DU(q,4)-\Delta(q,4)$, where
\begin{equation*}
\Delta(q,4)=
\begin{cases}
\left\lfloor q/4\right\rfloor&\text{if $q\equiv 0\pmod{3}$}, \\
0&\text{if $q\equiv 1\pmod{3}$}, \\
\left\lfloor (q-2)/6\right\rfloor&\text{if $q\equiv 2\pmod{3}$}
\end{cases}
\end{equation*}
for all $q$. \item A perfect $(5,3)_q\text{-DCC}$ of size
$DU(q,5)-\Delta(q,5)$, where
\begin{equation*}
\Delta(q,5) =
\begin{cases}
q/5&\text{if $q\equiv 0\pmod{10}$}, \\
0&\text{if $q\equiv 1\pmod{10}$}, \\
(3q-16)/10&\text{if $q\equiv 2\pmod{10}$},\\
(8q-14)/10&\text{if $q\equiv 3\pmod{10}$}, \\
(4q-6)/10&\text{if $q\equiv 4\pmod{10}$}, \\
0&\text{if $q\equiv 5\pmod{10}$}, \\
(q-16)/10&\text{if $q\equiv 6\pmod{10}$},\\
(6q-14)/10 &\text{if $q\equiv 9\pmod{10}$}
\end{cases}
\end{equation*}
for all $q\equiv 0,1,2,3,4,5,6$, or $9\pmod{10}$, except possibly
for $q\in\{13,14,15,16\}$.
\end{remunerate}
\end{theorem}

Recently, Wang \cite{Wang:2008} proved the
existence of two more classes of perfect $(n,n-2)_q\text{-DCCs}$
achieving optimality.

\begin{theorem}[Wang \cite{Wang:2008}]
\label{Bours}
The following codes exist.

\begin{remunerate}
\item A perfect $(4,2)_q\text{-DCC}$
of size $DU(q,4)-\Delta(q,4)$, where
\begin{equation*}
\Delta(q,4)=
\begin{cases}
 1&\text{if $q=9$}, \\
 0&\text{otherwise}
\end{cases}
\end{equation*}
for all $q$.

\item A perfect $(5,3)_q\text{-DCC}$ of size $DU(q,5)-\Delta(q,5)$, where
\begin{equation*}
\Delta(q,5) =
\begin{cases}
1&\text{if $q\equiv 7,9\pmod{10}$}, \\
0&\text{otherwise} \\
\end{cases}
\end{equation*}
for all $q$, except possibly for $q\in\{13,15,19,27,34\}$.
\end{remunerate}
\end{theorem}

Our interest in this paper is on the general problem of
determining the spectrum of possible sizes of a perfect
$(n,n-2)_q\text{-DCC}$. The main contributions of this paper are
the determination of

\begin{remunerate}
\item the spectrum of sizes of a perfect $(3,1)_q\text{-DCC}$ for
all $q$; and \item the spectrum of sizes of a perfect
$(4,2)_q\text{-DCC}$ for all but $19$ values of $q$.
\end{remunerate}

Our approach is design-theoretic and we begin by reviewing some notation and
terminology as well as required
concepts and prior results in combinatorial design theory.

\section{Mathematical preliminaries}
\label{sec3}

\subsection{Notation and terminology}
\label{sec3.1}

The ring of integers modulo $n$, $\bbZ/n\bbZ$, is denoted $\bbZ_n$.

For integers $m\leq n$, the set $\{m,m+1,\ldots,n\}$ is denoted
$[m,n]$, and we further abbreviate $[1,n]$ to $[n]$. Given a finite
set $X$ and an integer $k\in[|X|]$, we define the following:
\begin{align*}
\binom{X}{k} & = \{ A\subseteq X : |A|=k\}, \\
X^n & = \{ (x_1,\ldots,x_n):\text{$x_i\in X$ for all $i\in[n]$}\}, \\
X^n_* & = \{ (x_1,\ldots,x_n)\in X^n:\text{$x_i\not= x_j$, $1\leq i<j\leq n$}\}.
\end{align*}
Let $m\in[n]$.
A vector $(x_1,\ldots,x_n)\in X^n$ is said to {\em contain} a vector $(y_1,\ldots,y_m)\in X^m$ if
$y_i=x_{s_i}$ for some strictly increasing sequence of integers $(s_i)_{i\in[m]}$.

Given two sets $A,B\subseteq\bbZ$, $A+B$ denotes their {\em Minkowski sum}
$\{a+b:a\in A, b\in B\}$, and
for sets $A_i\subseteq\bbZ$, we
denote $A_1+\cdots+A_k$ by $\sum_{i=1}^k A_i$.

For $A=\{a_1,\ldots,a_k\}\subseteq\bbZ$,
$A_\uparrow$ denotes the vector whose components are the elements of $A$
sorted in ascending order, that is, $A_\uparrow=(a_1,\ldots,a_k)$, where
$a_1<\cdots< a_k$. If $A_\uparrow=(a_1,\ldots,a_k)$, then the vector
$(a_k,\ldots,a_1)$ is denoted $A_\downarrow$.

\subsection{Set systems}
\label{sec3.2}

A {\em set system} is a pair $(X,\A)$, where $X$ is a finite set of {\em points}
and $\A\subseteq 2^X$, whose elements are called {\em blocks}. The {\em order} of the
set system is $|X|$, the number of points.
For a set of nonnegative integers $K$, a set system $(X,\A)$ is said to be $K$-{\em uniform} if
$|A|\in K$ for all $A\in\A$.

Let $(X,\A)$ be a set system, and let $\B\subseteq\A$. If
$\sum_{B\in\B} |B|=|X|$ and $\cup_{B\in\B} B=X$, then $\B$ is
called a {\em parallel class} of $(X,\A)$. If $\sum_{B\in\B}
|B|=|X|-1$ and $\cup_{B\in\B} B=X\setminus\{x\}$ for some $x\in
X$, then $\B$ is called a {\em near-parallel class} of $(X,\A)$.
If $\A$ can be partitioned into parallel classes, then $(X,\A)$ is
said to be {\em resolvable}. If $\A$ can be partitioned into
near-parallel classes, then $(X,\A)$ is said to be {\em
near-resolvable}.

\subsection{Packings}
\label{sec3.3}

A {\em packing}, ${\rm P}(q,n,\lambda)$,
of order $q$, block size~$n$, and index $\lambda$ is a set system $(X,\A)$ of order $q$
such that the following conditions are satisfied:

\begin{enumerate}
\item[(a)]
$\A$ is a subset of elements of $\binom{X}{n}$, called {\em blocks}; and

\item[(b)]
every element of $\binom{X}{2}$ is contained in at most $\lambda$ blocks.
\end{enumerate}

\noindent
If we replace $\binom{X}{n}$ by $X^n$ in condition (a) and replace $\binom{X}{2}$ by $X^2$ in
condition~(b), we obtain
the definition of a {\em directed packing of order $q$, block size~$n$, and index $\lambda$}, denoted
${\rm DP}(q,n,\lambda)$. If we replace $\binom{X}{n}$ by $X^n_*$ in condition (a) and replace
$\binom{X}{2}$ by $X^2_*$ in condition~(b),
we have the definition of a {\em restricted directed packing of
order $q$, block size~$n$, and index $\lambda$}, denoted ${\rm RDP}(q,n,\lambda)$.
Throughout this paper, we adopt the convention that $\lambda$ is omitted from our notation
if $\lambda=1$.
A {\em perfect} ${\rm P}(q,n,\lambda)$ is one for which

\begin{enumerate}
\item[($\text{b}'$)] every element of $\binom{X}{2}$ is contained
in exactly $\lambda$ blocks.
\end{enumerate}

\noindent
{\em Perfect} ${\rm DP}(q,n,\lambda)$ and {\em perfect} ${\rm RDP}(q,n,\lambda)$
are similarly defined (by replacing $\binom{X}{2}$ in condition ($\text{b}'$)
with $X^2$ and $X^2_*$, respectively).

A perfect ${\rm P}(q,n,\lambda)$ is known as a {\em balanced incomplete block design} (BIBD)
and is denoted ${\rm B}(q,n,\lambda)$. A perfect ${\rm RDP}(q,n,\lambda)$ is known as a
{\em directed balanced incomplete block design} (DBIBD) and is denoted ${\rm DB}(q,n,\lambda)$.

\begin{theorem}[Hanani \cite{Hanani:1975}]
\label{Hanani}
Let $n\in\{4,5\}$ and $q\geq n$. Then
there exists a ${\rm B}(q,n,\lambda)$ if and only if $\lambda q(q-1)\equiv 0\pmod{n(n-1)}$
and $\lambda(q-1)\equiv 0\pmod{n-1}$, except for $(q,n,\lambda)=(15,5,2)$.
\end{theorem}

The maximum number of blocks in a ${\rm P}(q,n,\lambda)$,
${\rm DP}(q,n,\lambda)$, and ${\rm RDP}(q,n,\lambda)$ is denoted by
$D(q,n,\lambda)$, $DD(q,n, \lambda)$, and $RDD(q,n,\lambda)$, respectively. A ${\rm
P}(q,n,\lambda)$ with $D(q,n,\lambda)$ blocks, a ${\rm
DP}(q,n,\lambda)$ with $DD(q,n,\lambda)$ blocks, and an ${\rm
RDP}(q,n,\lambda)$ with $RDD(q,n,\lambda)$ blocks are called {\em
optimal}. Johnson \cite{Johnson:1962} and Sch\H{o}nheim
\cite{Schonheim:1966} showed that
\begin{equation*}
D(q,n,\lambda) \leq
\left\lfloor \frac{q}{n}\left\lfloor \frac{\lambda(q-1)}{n-1}\right\rfloor\right\rfloor =:
U(q,n,\lambda).
\end{equation*}
Using similar counting arguments, the following Johnson-type bound holds:
\begin{equation}
\label{johnsonRDD}
RDD(q,n,\lambda) \leq
\left\lfloor \frac{q}{n}\left\lfloor \frac{2\lambda(q-1)}{n-1}\right\rfloor\right\rfloor =:
RDU(q,n,\lambda).
\end{equation}

\begin{theorem}[Skillicorn \cite{Skillicorn:1982b}; Assaf, Shalaby, and Yin \cite{Assafetal:2001}]
\label{AssafShalabyYin} 
For all positive integers $q$,
$RDD(q,4)=RDU(q,4)$ except $RDD(9,4)=RDU(9,4)-1$.
\end{theorem}

Directed packings have featured strongly in the study of
$(n,n-2)_q\text{-DCCs}$ due to the following equivalence, for which we give a proof for the sake
of completeness.

\begin{proposition}[folklore]
\label{equivalence}
$\C\subseteq X^n$ is an $(n,n-2)_q\text{-DCC}$ if and only if $(X,\C)$ is a ${\rm DP}(q,n)$.
\end{proposition}

\begin{proof}
We need only check that every element of $X^2$ is contained in at
most one block in $\C$. Suppose this is not the case. Then there
are two blocks $\vecu,\vecv\in\C$ containing a common vector
$(a,b)\in X^2$. The Levenshte{\u\i}n distance between $\vecu$ and
$\vecv$ is therefore at most $2(n-2)$ since we can obtain $\vecv$
from $\vecu$ by deleting $n-2$ components from $\vecu$ leaving
$(a,b)$ and then inserting $n-2$ components to get $\vecv$. Hence
$\C$ is not $(n-2)$-deletion-correcting since any
$(n-2)$-deletion-correcting code must have a Levenshte{\u\i}n
distance of at least $2(n-2)+1$ between every pair of codewords.
\qquad\end{proof}

In view of Proposition~\ref{equivalence}, $\C$ can interchangeably be called an
$(n,n-2)_q\text{-DCC}$ or a ${\rm DP}(q,n)$. The following is also now evident.

\begin{proposition}
\label{prop3.4}
$\C\subseteq X^n$ is a perfect $(n,n-2)_q\text{-DCC}$ if and only if $(X,\C)$ is a perfect
${\rm DP}(q,n)$.
\end{proposition}

\begin{proof}
If $\C$ is a perfect $(n,n-2)_q\text{-DCC}$, then the balls
$D_{n-2}(\vecu)$, $\vecu\in\C$, partition $X^2$,
which is just another way of saying that every element of $X^2$ is contained in
exactly one block.
\qquad\end{proof}

Restricted directed packings do not directly
give useful deletion-correcting codes for real-world communication channels since the constraint
of not allowing repeated components in codewords is seldom imposed by such channels.
For example, in a noisy binary channel, it is necessary for codewords to contain repeated components
for any meaningful communication to take place. However, restricted directed packings
are important because they can be used as subcodes of deletion-correcting codes, allowing
repeated components, and, moreover, an optimal restricted directed packing can be used to construct
an optimal directed packing.

\begin{lemma}
\label{lem3.5}
There exists an ${\rm RDP}(q,n)$ having $M$ blocks if and only if there exists a
${\rm DP}(q,n)$ having $M+q$ blocks. Furthermore,
if the ${\rm RDP}(q,n)$ is a ${\rm DB}(q,n)$,
then the ${\rm DP}(q,n)$ is perfect and optimal.
\end{lemma}

\begin{proof}
If $(X,\A)$ is an ${\rm RDP}(q,n)$ with $M$ blocks, then $(X,\A\cup\{(x,\ldots,x)\in X^n\})$ is a
${\rm DP}(q,n)$ with $M+q$ blocks. It is easy to check that this ${\rm DP}(q,n)$ is perfect if $(X,\A)$
is a ${\rm DB}(q,n)$. To show that this perfect ${\rm DP}(q,n)$ is optimal, note that
$M=RDU(q,n)$ if $(X,\A)$ is a ${\rm DB}(q,n)$.

Conversely, if $(X,\A)$ is a ${\rm DP}(q,n)$ with $M+q$ blocks,
then deleting all the blocks containing the pairs $(x,x)\in X^2$
gives an ${\rm RDP}(q,n)$ with at least $M$ blocks.
\qquad\end{proof}

The following are two consequences of the above lemma.

\begin{corollary}
\label{DD=RDD+q}
$DD(q,n)=RDD(q,n)+q$.
\end{corollary}

\begin{corollary}
\label{DB}
For $n\in\{3,4,5,6\}$,
there exists a ${\rm DP}(q,n)$ that is perfect and optimal whenever $2q(q-1)\equiv 0\pmod{n(n-1)}$
and $2(q-1)\equiv 0\pmod{n-1}$, except for $(q,n)\in\{(15,5),(21,6)\}$.
\end{corollary}

\begin{proof}
The proof follows from the existence of ${\rm DB}(q,n)$ for $n\in\{3,4,5,6\}$
\cite{Bennettetal:1993,HungMendelsohn:1973,StreetSeberry:1980,StreetWilson:1980}.\qquad\end{proof}

The {\em leave graph} of a ${\rm DP}(q,n,\lambda)$ $(X,\A)$ is a directed loopless multigraph
$\Gamma$ with $V(\Gamma)=X$, and
an edge $(a,b)\in X^2_*$ appears $\lambda-s$ times in $E(\Gamma)$
if and only if $(a,b)$ appears in $s$ blocks of $\A$.

\subsection{Pairwise balanced designs and group divisible designs}
\label{sec3.4}

A {\em pairwise balanced design} (PBD) with set of block sizes $K$ is a
$K$-uniform set system $(X,\A)$,
such that every element of $\binom{X}{2}$ is contained in exactly one block of $\A$.
A PBD of order $q$ and set of block sizes $K$ is denoted ${\rm PBD}(q,K)$. If an element
$k\in K$ is ``starred'' (written $k^\star$), it means that the PBD has exactly one block of size~$k$.

\begin{theorem}[Brouwer \cite{Brouwer:1979}]
\label{47*} 
A ${\rm PBD}(q,\{4,7^\star\})$ exists if and only if
$q\equiv 7$ or $10\pmod{12}$, and $q\geq 22$.
\end{theorem}

Let $\lambda$ be a positive integer and $K\subseteq\bbZ$.
A {\em group divisible design} (GDD) {\em with set of block sizes $K$
and index $\lambda$}, denoted $(K,\lambda)\text{-GDD}$,
is a triple $(X,\G,\A)$, where

\begin{enumerate}
\item[(c)] 
$\A$ is a set of elements from $\cup_{k\in K}
\binom{X}{k}$, called {\em blocks};

\item[(d)]
$\G=\{G_1,\ldots,G_s\}$ is a partition of $X$ into subsets, called
{\em groups};

\item[(e)] 
every element of $\binom{X}{2}$ not
contained in a group is contained in exactly $\lambda$ blocks, and

\item[(f)] 
no block contains more than one point from any group.
\end{enumerate}

\noindent
The {\em type} of a GDD $(X,\G,\A)$ is the multiset $\{
|G|:G\in\G\}$. When it is more convenient, we use the exponential
notation to describe the type of a GDD: a GDD of type
$g_1^{t_1}\cdots g_s^{t_s}$ is a GDD where there are exactly $t_i$
groups of cardinality $g_i$, $i\in[s]$.

If we replace $\cup_{k\in K}\binom{X}{k}$ by $\cup_{k\in K} X^k_*$
in condition~(c) and replace $\binom{X}{2}$ by $X^2_*$ in
condition~(e), we get the definition of a {\em directed group
divisible design} (DGDD) {\em with set of block sizes $K$ and
index $\lambda$}, which we denote by $(K,\lambda)\text{-DGDD}$.

Again, we omit $\lambda$ from our notation if $\lambda=1$. Hence, we write $K\text{-GDD}$ and
$K\text{-DGDD}$ instead of $(K,1)\text{-GDD}$ and $(K,1)\text{-DGDD}$.

A $\{k\}\text{-GDD}$ of type $n^k$ is called a {\em transversal design} and is denoted ${\rm TD}(k,n)$.

A $(K,\lambda)\text{-GDD}$ $(X,\G,\A)$ is said to be {\em resolvable}
if $(X,\A)$ is resolvable.

We have the following simple observation.

\begin{lemma}
\label{GDDtoDGDD}
If there exists a $(K,\lambda)\text{-GDD}$ of type $g_1^{t_1}\cdots g_s^{t_s}$, then
there also exists a $(K,\lambda)\text{-DGDD}$ of type $g_1^{t_1}\cdots g_s^{t_s}$.
\end{lemma}

\begin{proof}
If $(X,\G,\A)$ is a $(K,\lambda)\text{-GDD}$ of type $g_1^{t_1}\cdots g_s^{t_s}$,
then $(X,\G,\B)$, where
$\B=\{A_\uparrow, A_\downarrow: A\in\A\}$ is a
$(K,\lambda)\text{-DGDD}$ of type $g_1^{t_1}\cdots g_s^{t_s}$.
\qquad\end{proof}

Another useful construction for DGDDs from GDDs is the following
``averaging construction.''

\begin{lemma}[averaging construction]
\label{average} 
Let $x\equiv y$ {\rm (mod~2)}, $x\geq y$. If there
exist $K$-{\rm GDD}s of types $g^n x^1$ and $g^n y^1$, then there
exists a $K$-{\rm DGDD} of type $g^n (\frac{x+y}{2})^1$.
\end{lemma}

\begin{proof}
Let $V=[gn]$, and let $\{G_1,\ldots,G_n\}$ be a partition of $V$
into $n$ parts of size~$g$. Let $X=\{a_1,\ldots,a_x\}$ be disjoint
from $V$, and let $Y=\{a_1,\ldots,a_y\}$,
$Z=\{a_1,\ldots,a_{(x+y)/2}\}$. Suppose that $(V\cup
X,\{G_1,\ldots,G_n,X\},\A)$ is a $K$-GDD of type $g^n x^1$ and
$(V\cup Y,\{G_1,\ldots,G_n,Y\},\B)$ is a $K$-GDD of type $g^n
y^1$.

For each block $A\in\A$ such that $a_i\in A$ and $i\in[(x+y)/2]$,
let $T_1(A)$ be the ordered tuple $(a_i,u_1,\ldots,u_k)$ such that
$\{a_i,u_1,\ldots,u_k\}=A$ and $u_1<\cdots<u_k$. For each block
$A\in\A$ such that $a_i\in A$ and $i\in[(x+y)/2+1,x]$, let
$T_2(A)$ be the ordered tuple $(u_1,\ldots,u_k,a_i)$ such that
$\{a_i,u_1,\ldots,u_k\}=A$ and $u_1<\cdots<u_k$. For each block
$A\in\A$ such that $a_i\not\in A$ for all $i\in[x]$, let
$T_3(A)=A_\uparrow$.

For each block $B\in\B$ such that $a_i\in B$ and $i\in[y]$, let
$T_4(B)$ be the ordered tuple $(u_1,\ldots,u_k,a_i)$ such that
$\{a_i,u_1,\ldots,u_k\}=B$ and $u_1>\cdots>u_k$. For each block
$B\in\B$ such that $a_i\not\in B$ for all $i\in[y]$, let
$T_5(B)=B_\downarrow$.

Define
\begin{equation*}
\C = \{T_1(A), T_2(A), T_3(A) : A\in \A\} \cup \{T_4(B),T_5(B): B\in\B\}.
\end{equation*}

Now, for each block in $\C$ that contains $a_i$, $i\in[x]$, relabel
the point $a_{x+1-j}$ as $a_{y+j}$, $j\in[(x-y)/2]$, and denote the
resulting blocks as $\D$. Then $(V\cup Z, \{G_1,\ldots,G_n,Z\},
\D)$ is a $K$-DGDD of type $g^n (\frac{x+y}{2})^1$.
\qquad\end{proof}

The following results on the existence of TDs, GDDs, and DGDDs are needed.

\begin{theorem}[folklore; see \cite{Abeletal:1996}]
\label{TD}
There exists a ${\rm TD}(k,n)$ for all $k\leq n+1$ whenever $n$ is a prime power.
\end{theorem}

\begin{theorem}[Zhu \cite{Zhu:1993}]
\label{3GDD} 
The necessary and sufficient conditions for the
existence of a $(\{3\},\lambda)\text{-GDD}$ of type $m^u$ are

\begin{remunerate}
\item $u \ge 3$; \item $\lambda (u-1)m \equiv 0\pmod{2}$; and

\item $\lambda u(u-1)m^2 \equiv 0\pmod{6}$.
\end{remunerate}
\end{theorem}

\begin{theorem}[Assaf and Hartman \cite{AssafHartman:1989}; Rees and Stinson \cite{ReesStinson:1987}; Rees \cite{Rees:1993}]
\label{3RGDD}
A resolvable $(\{3\},\lambda)\text{-GDD}$ of type $g^{v/g}$ exists
if and only if $\lambda(v-g)\equiv 0\pmod{2}$, $v\equiv
0\pmod{3}$, and $(\lambda,g,v)\notin\{(1,2,12),
(1,6,18)\}\cup\{(2j+1,2,6),(4j+2,1,6):j\geq 0\}$.
\end{theorem}

\begin{theorem}[Colbourn, Hoffman, and Rees \cite{Colbournetal:1992b}]
\label{CHR}
Let $g$, $t$, and $u$ be nonnegative integers. There exists a $\{3\}\text{-GDD}$ of type
$g^tu^1$ if and only if all of the following conditions are satisfied:

\begin{remunerate}
\item if $g>0$, then $t\geq 3$, or $t=2$ and $u=g$, or $t=1$ and $u=0$, or $t=0$;

\item $u\leq g(t-1)$ or $gt=0$;

\item $g(t-1)+u\equiv 0\pmod{2}$ or $gt=0$;

\item $gt\equiv 0\pmod{2}$ or $u=0$;

\item $g^2\binom{t}{2}+gtu\equiv 0\pmod{3}$.
\end{remunerate}
\end{theorem}

\begin{theorem}[Brouwer, Schrijver, and Hanani \cite{Brouweretal:1977}]
\label{4GDDg^t}
There exists a $\{4\}\text{-GDD}$ of type $g^t$ if and only if $t\geq 4$ and

\begin{remunerate}
\item $g\equiv 1$ or $5\pmod{6}$ and $t\equiv 1$ or $4\pmod{12}$; or

\item $g\equiv 2$ or $4\pmod{6}$ and $t\equiv 1\pmod{3}$; or

\item $g\equiv 3\pmod{6}$ and $t\equiv 0$ or $1\pmod{4}$; or

\item $g\equiv 0\pmod{6}$,
\end{remunerate}

\noindent
with the two exceptions of types $2^4$ and $6^4$, for which $\{4\}\text{-GDDs}$ do not exist.
\end{theorem}

\begin{theorem}[Brouwer \cite{Brouwer:1979}]
\label{4GDD2^t5^1}
There exists a $\{4\}\text{-GDD}$ of type $2^t5^1$
if and only if $t\equiv 0\pmod{3}$, and $t\geq 9$.
\end{theorem}

\begin{theorem}[Ge and Ling \cite{GeLing:2004}]
\label{4GDD12^um^1} 
There exists a $\{4\}\text{-GDD}$ of type
$12^u m^1$ for every $u\geq 4$ and $m\equiv 0\pmod{3}$, $0\leq
m\leq 6(u-1)$.
\end{theorem}

\begin{theorem}[Ge and Rees \cite{GeRees:2004}]
\label{4GDD6^um^1}
There exists a $\{4\}\text{-GDD}$ of type $6^u m^1$ for every $u\geq 4$
and $m\equiv 0\pmod{3}$, with $0\leq m\leq 3u-3$, except for
$(u,m)=(4,0)$ and except possibly for $(u,m)\in\{(7,15)$, $(11,21)$, $(11,24)$, $(11,27)$, $(13,27)$,
$(13,33)$, $(17,39)$, $(17,42)$, $(19,45)$, $(19,48)$, $(19,51)$, $(23,6)$, $(23,63)\}$.
\end{theorem}

\begin{theorem}[Sarvate \cite{Sarvate:1985}]
\label{4DGDD}
There exists a $\{4\}\text{-DGDD}$ of type $g^{v/g}$ if and only if $v\geq 4g$,
$v\equiv 0\pmod{g}$,
$v\equiv g\pmod{3}$, and $v(v-g)\equiv 0\pmod{6}$.
\end{theorem}

\subsection{Incomplete objects}
\label{sec3.5}

A {\em perfect} IDP {\em of order
$q$ and block size~$n$, with a hole of size~$h$}, denoted {\em
perfect ${\rm IDP}(q:h,n)$}, is a triple $(X,Y,\A)$, where
$(X,\A)$ is a ${\rm DP}(q,n)$ and $Y$ is an $h$-subset of $X$ such
that

\begin{remunerate}
\item no element of $Y^2$ is contained in any block of $\A$; and

\item every element of $X^2\setminus Y^2$ is contained in exactly one block of $\A$.
\end{remunerate}

Note that a perfect ${\rm IDP}(q:0,n)$ is equivalent to a perfect
${\rm DP}(q,n)$.

\subsection{Directability}
\label{sec3.6}

Suppose one is given an ${\rm RDP}(q,n,\lambda)$ $(X,\A)$ and we
replace each block $(a_1,\ldots,a_n)\in\A$ by the $n$-subset
$\{a_1,\ldots,a_n\}$. Then the resulting collection of $n$-subsets
contains every element of $\binom{X}{2}$ at most $2\lambda$ times.
This produces a ${\rm P}(q,n,2\lambda)$, the {\em underlying
packing} of the ${\rm RDP}(q,n,\lambda)$. A ${\rm
P}(q,n,2\lambda)$ is {\em directable} if it is the underlying
packing of some ${\rm RDP}(q,n,\lambda)$.

Harms and Colbourn \cite{HarmsColbourn:1983,HarmsColbourn:1985} developed an algorithm for directing ${\rm B}(q,3,2\lambda)$ that has the following important consequence.

\begin{theorem}[see \cite{ColbournRosa:1992}]
\label{directable}
Every packing ${\rm P}(q,3,2\lambda)$ is directable.
\end{theorem}

\section{Perfect incomplete directed packings of small order}
\label{sec4}

\enlargethispage*{6pt}
Here, we show the existence of some perfect IDPs
of small orders that are required for our constructions
in subsequent sections.

\begin{lemma}
\label{13:1}
There exists a perfect ${\rm IDP}(13:1,4)$.
\end{lemma}

\begin{proof}
Take the point set to be $\bbZ_{12}\cup\{\infty\}$. The orbits of the nine blocks
\begin{equation*}
\begin{array}{ccc}
(0,3,6,0) & (3,1,8,3) & (6,5,6,3) \\
(9,3,9,2) & (0,9,11,4) & (6,7,10,2) \\
(9,0,8,10) & (3,\infty,5,10) & (9,7,\infty,6) \\
\end{array}
\end{equation*}
under the action of the group $\langle\alpha\rangle$, where
\begin{equation*}
\alpha=(0 \ 1 \ 2)(3 \ 4 \ 5)(6 \ 7 \ 8)(9 \ 10 \ 11)(\infty),
\end{equation*}
together with the
two additional blocks, $(\infty,0,1,2)$ and $(0,1,2,\infty)$, give a perfect
${\rm IDP}(13:1,4)$, with $\{\infty\}$ as the hole of size one.
\qquad\end{proof}

\begin{lemma}
\label{28:4}
There exists a perfect ${\rm IDP}(28:4,4)$.
\end{lemma}

\begin{proof}
Take the point set to be $(\bbZ_{12}\times\{0,1\})\cup\{\infty_0$, $\infty_1$, $\infty_2$, $\infty_3\}$.
The orbits of the nine blocks
\begin{equation*}
\begin{array}{ccc}
((0,1),(3,1),(0,1),(5,0)) & ((0,0),(0,1),(1,1),(0,0)) &
((0,0),(1,0),(5,1),(11,0)) \\
((0,1),(6,1),(5,1),(3,0)) & ((0,0),(11,1),(7,0),(9,1)) & (\infty_0,(0,0),(5,0),(8,1)) \\
(\infty_0,(0,1),(1,0),(7,1)) & ((0,0),(7,1),(2,0),\infty_0) &
((0,1),(4,0),(2,1),\infty_0) \\
\end{array}
\end{equation*}
under the action of the group $\langle\alpha\rangle$, where
\begin{equation*}
\alpha(x) = \begin{cases}
(y+1,i) &\text{if $x=(y,i)\in \bbZ_{12} \times \{0,1\}$}, \\
\infty_{i+1\text{ (mod~3)}} &\text{if $x=\infty_i$, $i\in\{0,1,2\}$},
\end{cases}
\end{equation*}
together with the 22 additional blocks
\begin{equation*}
\begin{array}{cc}
(\infty_3,(0,0),(4,0),(8,0)) & ((8,0),(4,0),(0,0),\infty_3) \\
(\infty_3,(1,0),(5,0),(9,0)) & ((9,0),(5,0),(1,0),\infty_3) \\
(\infty_3,(2,0),(6,0),(10,0)) & ((10,0),(6,0),(2,0),\infty_3) \\
(\infty_3,(3,0),(7,0),(11,0)) & ((11,0),(7,0),(3,0),\infty_3) \\
(\infty_3,(0,1),(4,1),(8,1)) & ((8,1),(4,1),(0,1),\infty_3) \\
(\infty_3,(1,1),(5,1),(9,1)) & ((9,1),(5,1),(1,1),\infty_3) \\
(\infty_3,(2,1),(6,1),(10,1)) & ((10,1),(6,1),(2,1),\infty_3) \\
(\infty_3,(3,1),(7,1),(11,1)) & ((11,1),(7,1),(3,1),\infty_3) \\
((0,0),(3,0),(6,0),(9,0)) & ((9,0),(6,0),(3,0),(0,0)) \end{array}\end{equation*}\begin{equation*}\begin{array}{cc}
((1,0),(4,0),(7,0),(10,0)) & ((10,0),(7,0),(4,0),(1,0)) \\
((2,0),(5,0),(8,0),(11,0)) & ((11,0),(8,0),(5,0),(2,0)) \\
\end{array}
\end{equation*}
give a perfect ${\rm IDP}(28:4,4)$ with $\{\infty_0,\infty_1,\infty_2,\infty_3\}$ as the hole
of size four.
\qquad\end{proof}

\begin{lemma}
\label{40:4}
There exists a perfect ${\rm IDP}(40:4,4)$.
\end{lemma}

\begin{proof}
Take the point set to be $(\bbZ_{18} \times \{0,1\})\cup\{\infty_0$, $\infty_1$, $\infty_2$, $\infty_3\}$.
The orbits of the 13~blocks
\begin{equation*}
\begin{array}{cc}
((0,1),(11,1),(9,1),(15,0)) & ((0,1),(3,1),(0,1),(5,0)) \\
((0,1),(2,1),(7,1),(10,0)) & ((0,0),(2,0),(4,1),(5,0)) \\
((0,0),(4,0),(3,1),(15,0)) & ((0,1),(9,0),(4,1),(17,1)) \\
((0,0),(17,0),(11,1),(7,0)) & ((0,0),(0,1),(1,1),(0,0)) \\
((0,0),(16,1),(14,0),(6,1)) & ((0,0),(13,0),(9,1),\infty_0) \\
((1,1),(14,0),(11,1),\infty_0) & (\infty_0,(0,0),(5,1),(16,0)) \\
(\infty_0,(1,1),(8,0),(15,1)) \\
\end{array}
\end{equation*}
under the action of the group $\langle\alpha\rangle$, where
\begin{equation*}
\alpha(x) = \begin{cases}
(y+1,i)&\text{if $x=(y,i)\in \bbZ_{18} \times \{0,1\}$}, \\
\infty_{i+1 \text{ (mod~3)}} &\text{if $x=\infty_i$, $i\in\{0,1,2\}$},
\end{cases}
\end{equation*}
and the orbit of the block $((0,0),(9,0),(1,0),(10,0))$
under the action of $\langle\beta\rangle$, where
\begin{equation*}
\beta((y,i)) = (y+2,i),
\end{equation*}
together with the 24 blocks
\begin{equation*}
\begin{array}{ccc}
(\infty_3,(0,0),(6,0),(12,0)) & (\infty_3,(0,1),(6,1),(12,1)) & (\infty_3,(1,0),(7,0),(13,0)) \\
(\infty_3,(1,1),(7,1),(13,1)) & (\infty_3,(2,0),(8,0),(14,0)) & (\infty_3,(2,1),(8,1),(14,1)) \\
(\infty_3,(3,0),(9,0),(15,0)) & (\infty_3,(3,1),(9,1),(15,1)) & (\infty_3,(4,0),(10,0),(16,0)) \\
(\infty_3,(4,1),(10,1),(16,1)) & (\infty_3,(5,0),(11,0),(17,0)) & (\infty_3,(5,1),(11,1),(17,1)) \\
((12,0),(6,0),(0,0),\infty_3) & ((12,1),(6,1),(0,1),\infty_3) & ((13,0),(7,0),(1,0),\infty_3) \\
((13,1),(7,1),(1,1),\infty_3) & ((14,0),(8,0),(2,0),\infty_3) & ((14,1),(8,1),(2,1),\infty_3) \\
((15,0),(9,0),(3,0),\infty_3) & ((15,1),(9,1),(3,1),\infty_3) & ((16,0),(10,0),(4,0),\infty_3) \\
((16,1),(10,1),(4,1),\infty_3) & ((17,0),(11,0),(5,0),\infty_3) & ((17,1),(11,1),(5,1),\infty_3) \\
\end{array}
\end{equation*}
give a perfect ${\rm IDP}(40:4,4)$ with $\{\infty_0,\infty_1,\infty_2,\infty_3\}$
as the hole of size four.
\qquad\end{proof}

\section{\boldmath Bounds on sizes of perfect $(n-2)$-deletion-correcting codes}
\label{sec5}

In the remaining sections of this paper, we determine ${\rm
Spec}(q,n,n-2)$ for all $q$ when $n=3$, and for all $q$ with a
small finite number of possible exceptions when $n=4$. We begin by
establishing the extremal elements of ${\rm Spec}(q,n,n-2)$.
Define
\begin{equation*}
DL(q,n):=\left\lceil \frac{q}{n}\left\lceil \frac{2q}{n-1}\right\rceil\right\rceil.
\end{equation*}
Then the following holds.

\begin{proposition}
\label{prop5.1}
${\rm Spec}(q,n,n-2)\subseteq\left[DL(q,n),DU(q,n)\right]$ for all $q$.
\end{proposition}

\begin{proof}
The upper bound follows from Theorem~\ref{bound}.
To establish the lower bound, let $\C\subseteq X^n$ be a perfect $(n,n-2)_q\text{-DCC}$, and
let $r_x$ denote the number of codewords in $\C$ containing the point $x\in X$.
For any $x\in X$, there is exactly one codeword $\vecu\in\C$ containing the pair $(x,x)$.
The maximum number of distinct pairs of the form $(x,a)$ and $(a,x)$
that $\vecu$ can contain is $2n-3$, achieved when $\vecu=(x,a_1,\ldots,a_{n-2},x)$,
where $a_1,\ldots,a_{n-2}$ are all distinct and different from $x$. The other codewords in $\C$
containing $x$ each contains at most $n-1$ pairs either of the form $(x,a)$ or $(a,x)$.
Hence, $(n-1)(r_x-1)+2n-3\geq 2q-1$, which implies
$r_x+1\geq \left\lceil 2q/(n-1) \right\rceil$.
Hence,
\begin{equation*}
n|\C| = \sum_{x\in X} (r_x+1) \geq \sum_{x\in X} \left\lceil
\frac{2q}{n-1} \right\rceil = q\left\lceil \frac{2q}{n-1}
\right\rceil,
\end{equation*}
giving $|\C| \geq DL(q,n)$.
\qquad\end{proof}

Let
\begin{equation*}
I(q,n)=\left[ DL(q,n),DU(q,n)\right].
\end{equation*}
We show in subsequent sections that, in fact,
\begin{equation*}
{\rm Spec}(q,3,1) = I(q,3) ~\text{ and }~
{\rm Spec}(q,4,2) = I(q,4)
\end{equation*}
for almost all $q$, leaving only a small finite number of cases in doubt.

The following ``filling-in-groups'' construction is useful.

\begin{proposition}
\label{membership}
If an $\{n\}\text{-DGDD}$ of type $\{ g_1,\ldots,g_s\}$ exists, then
\begin{align}
\label{membershipinI} \\
\left\{\frac{2}{\binom{n}{2}}\left(\binom{{\sum_{i=1}^s
g_i}}{2}-\sum_{i=1}^s\binom{g_i}{2}\right)\right\} +\sum_{i=1}^s
{\rm Spec}(g_i,n,n-2)
\subseteq {\rm Spec}\left(\sum_{i=1}^s g_i,n,n-2\right). \nn
\end{align}
\end{proposition}
\unskip

\begin{proof}
Let $(X,\A)$ be an $\{n\}\text{-DGDD}$ of type $\{
g_1,\ldots,g_s\}$. We produce a perfect ${\rm DP}(\sum_{i=1}^s
g_i,n)$ as follows. For each group $G_i\in\G$ of size $g_i$, we
add the blocks of a perfect ${\rm DP}(g_i,n)$ $(G_i,\B_i)$ to
$\A$. By using perfect ${\rm DP}(g_i,n)$'s of different sizes, we
obtain (\ref{membershipinI}), noting that the first term in
(\ref{membershipinI}) is the quantity $|\A|$.
\qquad\end{proof}

With a proof similar to that of
Proposition~\ref{membership}, we have the following construction.

\begin{proposition}
\label{membership-1}
Let $h$ be a positive integer. If we have

\begin{remunerate}
\item an $\{n\}\text{-DGDD}$ of type $\{ g_1,\ldots,g_s\}$;

\item a perfect ${\rm IDP}(g_i+h:h,n)$ having $t_i$ blocks for $1
\le i \le s-1$; and \item a perfect ${\rm DP}(g_s+h,n)$ having
$t$ blocks,
\end{remunerate}

\noindent
then
\begin{equation*}
\left\{\frac{2}{\binom{n}{2}}\left(\binom{{\sum_{i=1}^s
g_i}}{2}-\sum_{i=1}^s\binom{g_i}{2}\right)\right\}+t
+\sum_{i=1}^{s-1} t_i
 \in {\rm Spec}\left(h+\sum_{i=1}^s
g_i,n,n-2\right).
\end{equation*}
\end{proposition}
\unskip

\section{\boldmath The spectrum ${\rm Spec}(q,3,1)$}
\label{sec6}

\begin{table}[h!]
\footnotesize
\caption{Some perfect $(3,1)_q$-{\rm DCCs} with small $q$.}
\label{smallspec}
\centering
\begin{tabular}{c | c | l }
\hline
$q$ & Code size & Codewords of a perfect $(3,1)_q$-DCC \\
\hline
\hline
1 & 1 & $(0,0,0)$ \\
\hline
2 & 2 & $(0,0,0)$ $(1,0,1)$ \\
\hline
3 & 3 & $(0,1,0)$ $(1,2,1)$ $(2,0,2)$ \\
\cline{2-3}
 & 4 & $(0,0,1)$ $(0,2,2)$ $(1,2,0)$ $(2,1,1)$ \\
\cline{2-3}
 & 5 & $(0,0,0)$ $(0,2,1)$ $(1,1,1)$ $(1,2,0)$ $(2,2,2)$ \\
\hline
4 & 6 & $(0,0,2)$ $(0,1,3)$ $(1,2,1)$ $(2,2,0)$ $(3,1,0)$ $(3,2,3)$ \\
\cline{2-3}
 & 7 & $(0,0,0)$ $(0,1,2)$ $(1,3,3)$ $(2,0,3)$ $(2,1,1)$ $(3,1,0)$ \\
 & & $(3,2,2)$ \\
\cline{2-3}
 & 8 & $(0,0,0)$ $(0,3,2)$ $(1,1,1)$ $(1,2,3)$ $(2,0,1)$ $(2,2,2)$ \\
 & & $(3,1,0)$ $(3,3,3)$ \\
\hline
5 & 9 & $(0,3,2)$ $(0,4,0)$ $(1,2,3)$ $(1,4,4)$ $(2,1,0)$ $(2,4,2)$ \\
 & & $(3,0,1)$ $(3,4,3)$ $(4,1,1)$ \\
\cline{2-3}
 & 10 & $(0,1,0)$ $(0,4,2)$ $(1,1,2)$ $(1,3,3)$ $(2,1,4)$ $(2,3,0)$ \\
 & & $(3,2,2)$ $(3,4,1)$ $(4,0,3)$ $(4,4,4)$ \\
\cline{2-3}
 & 11 & $(0,0,0)$ $(0,1,3)$ $(0,2,4)$ $(1,1,1)$ $(1,2,0)$ $(2,2,2)$ \\
 & & $(2,3,1)$ $(3,3,3)$ $(3,4,2)$ $(4,1,4)$ $(4,3,0)$ \\
\hline
6 & 12 & $(0,1,0)$ $(1,5,1)$ $(2,0,2)$ $(2,1,4)$ $(3,0,4)$ $(3,1,3)$ \\
 & & $(3,2,5)$ $(4,0,3)$ $(4,1,2)$ $(4,5,4)$ $(5,0,5)$ $(5,2,3)$ \\
\cline{2-3}
 & 13 & $(0,3,1)$ $(0,4,0)$ $(1,1,3)$ $(1,2,0)$ $(1,4,5)$ $(2,3,2)$ \\
 & & $(2,5,4)$ $(3,0,5)$ $(4,2,1)$ $(4,3,4)$ $(5,0,2)$ $(5,3,3)$ \\
 & & $(5,5,1)$ \\
\cline{2-3}
 & 14 & $(0,1,1)$ $(0,3,2)$ $(0,4,5)$ $(1,0,0)$ $(2,1,4)$ $(2,2,2)$ \\
 & & $(2,3,0)$ $(3,1,5)$ $(3,4,3)$ $(4,1,2)$ $(4,4,4)$ $(5,1,3)$ \\
 & & $(5,2,5)$ $(5,4,0)$ \\
\cline{2-3}
 & 15 & $(0,2,2)$ $(0,4,1)$ $(1,1,1)$ $(1,3,4)$ $(1,5,0)$ $(2,0,3)$ \\
 & & $(2,5,1)$ $(3,0,5)$ $(3,1,2)$ $(3,3,3)$ $(4,0,0)$ $(4,4,2)$ \\
 & & $(4,5,3)$ $(5,2,4)$ $(5,5,5)$ \\
\cline{2-3}
 & 16 & $(0,0,0)$ $(0,4,1)$ $(0,5,3)$ $(1,1,1)$ $(1,3,5)$ $(2,1,0)$ \\
 & & $(2,2,2)$ $(2,4,3)$ $(3,1,2)$ $(3,3,3)$ $(3,4,0)$ $(4,2,5)$ \\
 & & $(4,4,4)$ $(5,0,2)$ $(5,1,4)$ $(5,5,5)$ \\
\hline
8 & 22 & $(5,1,7)$ $(5,3,6)$ $(4,0,4)$ $(3,0,1)$ $(6,4,5)$ $(1,4,6)$ \\
 & & $(6,0,2)$ $(6,1,3)$ $(3,4,3)$ $(2,4,2)$ $(3,2,7)$ $(6,7,6)$ \\
 & & $(1,2,1)$ $(7,5,4)$ $(1,5,0)$ $(7,0,0)$ $(7,2,3)$ $(4,7,1)$ \\
 & & $(0,3,5)$ $(5,2,5)$ $(0,7,7)$ $(2,0,6)$ \\
\cline{2-3}
 & 23 & $(7,7,6)$ $(2,6,1)$ $(0,3,6)$ $(7,5,1)$ $(6,5,0)$ $(1,6,7)$ \\
 & & $(2,5,2)$ $(6,3,2)$ $(1,0,2)$ $(0,0,5)$ $(2,4,3)$ $(5,5,7)$ \\
 & & $(4,5,6)$ $(1,5,3)$ $(0,7,4)$ $(3,0,1)$ $(4,7,2)$ $(1,4,1)$ \\
 & & $(2,7,0)$ $(3,5,4)$ $(6,6,4)$ $(4,4,0)$ $(3,7,3)$ \\
\cline{2-3}
 & 24 & $(7,7,3)$ $(5,4,6)$ $(3,1,5)$ $(6,2,3)$ $(5,1,1)$ $(7,4,2)$ \\
 & & $(3,4,0)$ $(6,7,5)$ $(0,6,0)$ $(7,6,6)$ $(4,3,7)$ $(5,0,3)$ \\
 & & $(2,1,6)$ $(0,1,7)$ $(2,2,2)$ $(3,3,6)$ $(1,3,2)$ $(4,5,5)$ \\
 & & $(2,0,4)$ $(2,5,7)$ $(7,1,0)$ $(6,4,1)$ $(0,5,2)$ $(1,4,4)$ \\
\cline{2-3}
 & 25 & $(0,2,5)$ $(6,6,6)$ $(3,3,2)$ $(0,1,3)$ $(3,6,0)$ $(5,1,1)$ \\
 & & $(1,5,5)$ $(2,2,7)$ $(7,0,6)$ $(3,7,7)$ $(1,2,6)$ $(6,7,4)$ \\
 & & $(5,7,2)$ $(7,5,3)$ $(4,5,6)$ $(3,1,4)$ $(4,7,1)$ $(4,2,3)$ \\
 & & $(6,2,1)$ $(5,4,0)$ $(0,0,0)$ $(2,0,4)$ $(6,3,5)$ $(4,4,4)$ \\
 & & $(1,0,7)$ \\
\cline{2-3}
 & 26 & $(3,1,4)$ $(2,0,7)$ $(6,2,3)$ $(4,0,5)$ $(6,6,6)$ $(2,1,5)$ \\
 & & $(5,7,1)$ $(2,4,6)$ $(5,4,2)$ $(1,1,1)$ $(2,2,2)$ $(0,1,2)$ \\
 & & $(1,0,3)$ $(4,4,4)$ $(4,3,7)$ $(7,0,4)$ $(0,0,0)$ $(3,3,5)$ \\
 & & $(7,7,7)$ $(7,6,5)$ $(5,6,0)$ $(7,3,2)$ $(5,5,3)$ $(1,6,7)$ \\
 & & $(6,4,1)$ $(3,0,6)$ \\
\hline
\end{tabular}
\end{table}

We begin by determining $I(q,3)$ for some small $q$.

\begin{lemma}
\label{specsmall}
For $q\in\{1,2,3,4,5,6,8,11,14\}$, we have ${\rm Spec}(q,3,1)=I(q,3)$.
\end{lemma}

\begin{proof}
For $q\in\{1,2,3,4,5,6,8\}$,
perfect $(3,1)_q\text{-DCCs}$ having number of blocks in $I(q,3)$
are given in Table~\ref{smallspec}.

For $q=11$, apply Proposition~\ref{membership} to a $\{3\}\text{-DGDD}$ of type $5^1 1^6$,
which exists by Lemma~\ref{GDDtoDGDD} and Theorem~\ref{CHR},
to conclude that $\{45,46,47\}\subseteq {\rm Spec}(11,3,1)$. Membership of
41, 42, 43, and 44 in ${\rm Spec}(11,3,1)$ is given by Table~\ref{midspec}.

For $q=14$, apply Proposition~\ref{membership} to a
$\{3\}\text{-DGDD}$ of type $4^3 2^1$, which exists by Lemma~\ref{GDDtoDGDD} and Theorem~\ref{CHR}, to conclude that
$[68,74]\subseteq {\rm Spec}(14,3,1)$. Membership of 66 and 67 in
${\rm Spec}(14,3,1)$ is given by Table~\ref{midspec}.
\qquad\end{proof}

\begin{table*}
\footnotesize
\caption{Some perfect $(3,1)_q\text{-DCCs}$ with $q\in\{11,14\}$.}
\label{midspec}
\centering
\begin{tabular}{c | c | l }
\hline
$q$ & Code size & Codewords of a perfect $(3,1)_q$-DCC \\
\hline
\hline
11 & 41 & $(5,0,8)$ $(6,7,9)$ $(5,1,5)$ $(9,7,2)$ $(8,7,10)$ $(0,1,9)$ \\
 & & $(9,3,9)$ $(1,7,8)$ $(1,2,0)$ $(4,9,8)$ $(9,10,1)$ $(2,1,3)$ \\
 & & $(8,6,1)$ $(4,10,7)$ $(4,3,6)$ $(2,7,6)$ $(6,5,6)$ $(8,2,8)$ \\
 & & $(3,10,5)$ $(3,8,4)$ $(2,2,4)$ $(6,8,0)$ $(5,3,7)$ $(7,0,7)$ \\
 & & $(0,6,4)$ $(2,5,9)$ $(7,3,1)$ $(10,8,3)$ $(6,3,2)$ $(4,4,2)$ \\
 & & $(0,5,2)$ $(5,10,4)$ $(1,6,10)$ $(3,0,3)$ $(1,4,1)$ $(9,4,0)$ \\
 & & $(7,4,5)$ $(8,9,5)$ $(10,2,10)$ $(10,9,6)$ $(0,10,0)$\\
\cline{2-3}
 & 42 & $(0,9,2)$ $(9,8,3)$ $(7,9,5)$ $(3,0,3)$ $(5,4,3)$ $(2,3,4)$ \\
 & & $(6,3,7)$ $(5,1,9)$ $(3,1,5)$ $(7,8,0)$ $(5,8,10)$ $(9,0,7)$ \\
 & & $(3,8,2)$ $(10,1,3)$ $(10,10,7)$ $(3,10,9)$ $(9,4,4)$ $(4,2,5)$ \\
 & & $(9,6,10)$ $(10,5,6)$ $(8,1,8)$ $(0,1,10)$ $(8,5,7)$ $(7,4,10)$ \\
 & & $(0,6,0)$ $(6,5,2)$ $(8,6,9)$ $(1,6,1)$ $(2,6,8)$ $(6,4,6)$ \\
 & & $(1,2,7)$ $(4,7,1)$ $(4,0,8)$ $(2,2,10)$ $(10,2,0)$ $(7,7,2)$ \\
 & & $(4,9,9)$ $(5,0,5)$ $(10,8,4)$ $(2,9,1)$$(7,3,6)$ $(1,0,4)$ \\
 \cline{2-3}
& 43 & $(8,7,0)$ $(10,6,3)$ $(5,10,8)$ $(8,3,3)$ $(0,4,9)$ $(5,2,1)$ \\
& & $(6,1,1)$ $(0,0,5)$ $(4,7,2)$ $(2,9,0)$ $(5,4,6)$ $(10,1,5)$ \\
& & $(7,5,7)$ $(10,0,7)$ $(6,8,9)$ $(10,2,4)$ $(1,7,9)$ $(9,3,7)$ \\
& & $(9,5,9)$ $(3,0,6)$ $(0,1,8)$ $(7,3,1)$ $(2,6,6)$ $(4,1,0)$ \\
 & & $(0,2,3)$ $(3,8,2)$ $(2,8,5)$ $(7,8,6)$ $(8,8,10)$ $(0,10,10)$\\
 & & $(9,8,1)$ $(6,2,7)$ $(6,5,0)$ $(7,4,10)$ $(9,2,10)$ $(3,10,9)$ \\
 & & $(1,4,3)$ $(5,5,3)$ $(3,4,5)$ $(9,6,4)$ $(1,6,10)$ $(1,2,2)$\\
 & & $(4,8,4)$ \\
\cline{2-3}
& 44 & $(7,9,3)$ $(10,0,9)$ $(2,8,8)$ $(6,6,1)$ $(7,8,6)$ $(6,7,4)$ \\
& & $(7,10,2)$ $(7,1,1)$ $(0,7,7)$ $(0,3,10)$ $(6,3,0)$ $(3,1,5)$ \\
& & $(8,2,5)$ $(4,7,5)$ $(10,4,4)$ $(5,7,0)$ $(2,10,6)$ $(9,4,0)$ \\
& & $(5,10,10)$ $(4,8,10)$ $(2,2,3)$ $(8,3,9)$ $(0,5,8)$ $(2,1,0)$ \\
 & & $(0,4,6)$ $(4,3,2)$ $(3,3,4)$ $(9,2,7)$ $(9,8,1)$ $(1,3,6)$ \\
 & & $(2,4,9)$ $(0,1,2)$ $(5,4,1)$ $(1,9,10)$ $(1,8,4)$ $(10,5,3)$\\
 & & $(9,5,5)$ $(6,10,8)$ $(10,1,7)$ $(3,8,7)$ $(8,0,0)$ $(6,9,9)$ \\
 & & $(6,5,2)$ $(5,9,6)$ \\
\hline
14 & 66 & $(4,7,6)$ $(8,2,1)$ $(4,12,1)$ $(2,5,9)$ $(13,5,7)$ $(0,6,11)$ \\
 & & $(7,4,11)$ $(12,0,10)$ $(6,2,6)$ $(10,12,6)$ $(1,7,8)$ $(7,13,1)$ \\
 & & $(9,8,10)$ $(11,10,13)$ $(9,5,1)$ $(5,6,4)$ $(2,0,3)$ $(10,8,7)$ \\
 & & $(5,0,5)$ $(0,9,2)$ $(1,4,10)$ $(2,12,2)$ $(8,13,11)$ $(5,11,8)$ \\
 & & $(11,6,7)$ $(13,9,12)$ $(3,1,1)$ $(2,11,4)$ $(9,13,4)$ $(2,7,10)$ \\
 & & $(3,0,12)$ $(13,10,2)$ $(12,4,13)$ $(6,0,8)$ $(1,3,3)$ $(4,9,0)$\\
 & &$(0,1,0)$ $(8,12,8)$ $(10,1,9)$ $(11,3,11)$ $(1,11,5)$ $(6,13,3)$ \\
 & &$(11,1,2)$ $(3,7,2)$ $(9,7,3)$ $(13,0,13)$ $(5,2,13)$ $(12,7,9)$ \\
 & &$(1,13,6)$ $(10,3,10)$ $(3,13,8)$ $(3,6,9)$ $(12,11,12)$ $(10,4,5)$ \\
 & & $(6,1,12)$ $(6,5,10)$ $(8,5,3)$ $(10,11,0)$ $(8,0,4)$ $(4,3,4)$ \\
 & & $(12,3,5)$ $(7,0,7)$ $(7,5,12)$ $(8,9,6)$ $(4,2,8)$ $(9,11,9)$ \\
\cline{2-3}
& 67 & $(6,5,10)$ $(8,0,8)$ $(12,5,3)$ $(13,3,3)$ $(13,10,11)$ $(3,10,10)$ \\
& & $(6,12,6)$ $(10,2,6)$ $(6,13,9)$ $(11,3,5)$ $(9,4,10)$ $(8,11,7)$ \\
& & $(8,13,5)$ $(12,0,2)$ $(10,4,4)$ $(3,8,2)$ $(4,8,12)$ $(9,7,12)$ \\
& & $(2,8,9)$ $(4,9,2)$ $(13,6,7)$ $(10,13,8)$ $(9,0,13)$ $(3,12,4)$ \\
& & $(5,4,13)$ $(5,0,6)$ $(10,5,12)$ $(0,4,3)$ $(11,6,2)$ $(2,3,1)$\\
& & $(11,8,1)$ $(1,10,3)$ $(9,6,11)$ $(5,9,5)$ $(8,6,3)$ $(4,1,6)$ \\
 & &$(0,1,5)$ $(12,9,1)$ $(12,8,10)$ $(7,4,11)$ $(11,0,11)$ $(11,10,9)$ \\
 & &$(5,7,8)$ $(7,3,7)$ $(12,11,12)$ $(0,7,9)$ $(1,0,12)$ $(1,13,1)$ \\
 & & $(5,1,2)$ $(6,8,4)$ $(13,4,0)$$(3,11,13)$ $(0,10,0)$ $(13,13,2)$ \\
 & & $(2,7,0)$ $(1,11,4)$ $(10,1,7)$ $(9,3,9)$ $(2,5,11)$ $(4,7,5)$ \\
 & & $(1,9,8)$ $(7,2,10)$ $(3,6,0)$ $(12,7,13)$ $(2,2,4)$ $(7,6,1)$ \\
 & &$(2,13,12)$ \\
\hline
\end{tabular}
\end{table*}

\begin{lemma}
\label{n=3,0,1(3)}
${\rm Spec}(q,3,1)=I(q,3)$ for all $q\equiv 0$ or $1\pmod{3}$.
\end{lemma}

\begin{proof}
The case $q\leq 6$ is settled by Lemma~\ref{specsmall}. If
$q\equiv 0\pmod{3}$, $q\geq 9$, apply Proposition~\ref{membership}
to a $\{3\}\text{-DGDD}$ of type $3^{q/3}$, obtained by directing
a $(\{3\},2)\text{-GDD}$ of type $3^{q/3}$, which exists by
Theorem~\ref{3GDD}, to conclude that
\begin{align*}
{\rm Spec}(q,3,1) & \supseteq \left\{\frac{q(q-3)}{3}\right\}+\sum_{i=1}^{q/3} {\rm Spec}(3,3,1) \\
& = \left[\frac{q^2}{3}, \frac{q^2+2q}{3}\right] \\
& = I(q,3).
\end{align*}

If $q\equiv 1\pmod{3}$, apply Proposition~\ref{membership} to a
$\{3\}\text{-DGDD}$ of type $3^{(q-1)/3}1^1$, obtained by
directing a $(\{3\},2)\text{-GDD}$ of type $3^{(q-1)/3}1^1$, to
conclude that
\begin{align*}
{\rm Spec}(q,3,1) &
\supseteq \left\{\frac{(q-1)(q-2)}{3}+1\right\}+
\sum_{i=1}^{(q-1)/3} {\rm Spec}(3,3,1) \\
 & = \left[ \frac{q^2+2}{3}, \frac{q^2+2q}{3}\right] \\
 & = I(q,3).
\end{align*}
A $(\{3\},2)\text{-GDD}$ of type $3^{(q-1)/3}1^1$
can be constructed by taking a near-resolvable ${\rm B}(q,3,2)$,
whose existence is well established (see, for example, \cite{Furinoetal:1996}), and considering one of its
near-parallel class as the set of groups.
\qquad\end{proof}

\begin{lemma}
\label{n=3,2(3)}
${\rm Spec}(q,3,1)=I(q,3)$ for all $q\equiv 2\pmod{3}$.
\end{lemma}

{\it Proof}.
The case $q\in\{8,11,14\}$ is settled by Lemma~\ref{specsmall}. If
$q\equiv 2\pmod{6}$, $q\geq 20$, apply Proposition~\ref{membership} to a $\{3\}\text{-DGDD}$ of type
$6^{(q-2)/6}2^1$, which exists by Lemma~\ref{GDDtoDGDD} and
Theorem~\ref{CHR}, to conclude that
\begin{align*}
{\rm Spec}(q,3,1) & \supseteq \left\{\frac{(q-2)(q-4)}{3}\right\} +
 \sum_{i=1}^{(q-2)/6} {\rm Spec}(6,3,1) + {\rm Spec}(2,3,1) \\
& = \left[ \frac{q^2+2}{3}, \frac{q^2+2q-2}{3}\right] \\
& = I(q,3).
\end{align*}

If $q\equiv 5\pmod{6}$, $q\geq 17$, apply Proposition~\ref{membership} to a $\{3\}\text{-DGDD}$ of type
$3^{(q-5)/3}5^1$, which exists by Lemma~\ref{GDDtoDGDD} and
Theorem~\ref{CHR}, to conclude that
\begin{align*}
{\rm Spec}(q,3,1) & \supseteq \left\{\frac{(q+2)(q-5)}{3}\right\}+
\sum_{i=1}^{(q-5)/3} {\rm Spec}(3,3,1)+ {\rm Spec}(5,3,1) \\
& = \left[ \frac{q^2+2}{3},\frac{q^2+2q-2}{3}\right] \\
& = I(q,3).
\qquad\endproof
\end{align*}

Lemmas~\ref{n=3,0,1(3)} and \ref{n=3,2(3)} combine to give the following.

\begin{theorem}
\label{thm6.4}
${\rm Spec}(q,3,1)=I(q,3)$ for all $q$.
\end{theorem}

\pagebreak
\section{\boldmath The spectrum Spec$(q,4,2)$}
\label{sec7}

\subsection{Nonexistence}
\label{sec7.1}

We begin with some nonexistence results.

\begin{lemma}
\label{non-existence}
The following deletion-correcting codes do not exist:

\begin{remunerate}
\item Perfect $(4,2)_4\text{-DCCs}$ of size $DL(4,4)=3$; and

\item perfect $(4,2)_6\text{-DCCs}$ of size $DL(6,4)+1=7$.
\end{remunerate}
\end{lemma}

\begin{proof}
The proof is established by exhaustive search.
\qquad\end{proof}

Next, we determine some elements of ${\rm Spec}(q,4,2)$ for small $q$, which are
required in our recursive constructions later.

\subsection{Small spectrum members}\
\label{sec7.2}

\begin{lemma}
\label{smallspec4lemma} 
For $q\in\{1,2,3,5,7\}$, we have ${\rm
Spec}(q,4,2)=I(q,4)$, and

\begin{remunerate}
\item ${\rm Spec}(4,4,2)=I(4,4)\setminus\{3\}$;

\item ${\rm Spec}(6,4,2)=I(6,4)\setminus\{7\}$.
\end{remunerate}
\end{lemma}

\begin{proof}
For $q\in[7]$, the existence of perfect $(4,2)_q\text{-DCCs}$ of the required sizes in
${\rm Spec}(q,4,2)$ are given in Table~\ref{smallspec4}.
\qquad\end{proof}

\begin{table}
\footnotesize
\caption{Some perfect $(4,2)_q\text{-DCCs}$ with small $q$.}
\label{smallspec4}
\centering
\begin{tabular}{c | c | l }
\hline
$q$ & Code & Codewords of a perfect $(4,2)_q\text{-DCC}$ \\
 & size & \\
\hline
\hline
1 & 1 & $(0,0,0,0)$ \\
\hline
2 & 1 & $(0,1,1,0)$ \\
\cline{2-3}
 & 2 & Exists by Theorem~\ref{Bours} \\
\hline
3 & 2 & $(0,2,1,0)$ $(1,1,2,2)$ \\
\cline{2-3}
 & 3 & Exists by Theorem~\ref{Bours} \\
\hline
4 & 3 & Does not exist by Lemma~\ref{non-existence} \\
\cline{2-3}
& 4 & $(0,2,0,1)$ $(1,3,0,3)$ $(2,2,2,3)$ $(3,1,1,2)$ \\
\cline{2-3}
 & 5 & $(0,0,0,2)$ $(0,1,1,1)$ $(0,3,3,3)$ $(2,1,3,0)$ $(3,1,2,2)$ \\
\cline{2-3}
 & 6 & Exists by Theorem~\ref{Bours} \\
\hline
5 & 5 & $(0,1,0,2)$ $(1,4,1,3)$ $(2,3,2,1)$ $(3,0,3,4)$ $(4,2,4,0)$ \\
\cline{2-3}
 & 6 & $(0,4,2,3)$ $(1,1,1,3)$ $(1,2,0,0)$ $(2,1,4,4)$ $(3,3,2,2)$ \\
 & & $(3,4,0,1)$ \\
\cline{2-3}
 & 7 & Exists by Theorem~\ref{Bours} \\
\hline
6 & 6 & $(1,2,0,1)$ $(3,1,4,3)$ $(4,0,2,4)$ $(5,4,1,5)$ $(0,3,5,0)$ \\
 & & $(2,5,3,2)$ \\
\cline{2-3}
 & 7 & Does not exist by Lemma~\ref{non-existence} \\
\cline{2-3}
 & 8 & $(4,2,5,5)$ $(5,2,0,3)$ $(0,2,4,1)$ $(1,3,0,5)$ $(3,1,2,2)$ \\
 & & $(4,0,0,0)$ $(3,4,4,3)$ $(5,1,1,4)$ \\
\cline{2-3}
 & 9 & Exists by Theorem~\ref{Bours-1} \\
\cline{2-3}
 & 10 & Exists by Theorem~\ref{Bours} \\
\hline
7 & 9 & $(0,1,2,0)$ $(1,3,4,1)$ $(0,3,5,3)$ $(2,3,6,2)$ $(4,2,4,5)$ \\
 & & $(4,6,3,0)$ $(5,0,6,4)$ $(5,5,2,1)$ $(6,1,6,5)$ \\
\cline{2-3}
 & 10 & $(6,4,1,5)$ $(3,3,5,6)$ $(4,2,4,3)$ $(0,4,0,6)$ $(1,0,1,3)$ \\
 & & $(6,3,0,2)$ $(2,1,2,6)$ $(5,3,1,4)$ $(5,2,0,5)$ $(6,6,6,6)$ \\
\cline{2-3}
 & 11 & $(6,3,6,5)$ $(4,6,1,0)$ $(3,0,3,2)$ $(5,5,5,4)$ $(1,5,6,2)$ \\
 & & $(4,2,5,3)$ $(2,0,6,4)$ $(1,3,1,4)$ $(0,5,0,1)$ $(2,2,2,1)$ \\
 & & $(4,4,4,4)$ \\
\cline{2-3}
 & 12 & $(3,2,1,0)$ $(6,6,6,1)$ $(0,0,0,3)$ $(1,1,5,2)$ $(0,2,4,5)$ \\
 & & $(4,0,1,6)$ $(5,5,6,0)$ $(3,3,6,5)$ $(5,1,3,4)$ $(6,4,2,3)$ \\
 & & $(2,2,2,6)$ $(4,4,4,4)$ \\
\cline{2-3}
 & 13 & $(6,6,5,6)$ $(0,6,4,2)$ $(3,3,0,3)$ $(6,0,0,0)$ $(2,3,6,1)$ \\
 & & $(4,0,5,1)$ $(1,1,1,1)$ $(4,4,4,4)$ $(5,5,5,3)$ $(1,2,5,0)$ \\
 & & $(2,2,2,2)$ $(3,5,2,4)$ $(1,4,6,3)$ \\
\cline{2-3}
& 14 & Exists by Theorem~\ref{Bours} \\
\hline
\end{tabular}
\end{table}

\begin{lemma}
\label{DL for 10} 
$DL(10,4)=18\in {\rm Spec}(10,4,2)$.
\end{lemma}

\begin{proof}
Take the alphabet $X$ to be $\{0,1,\ldots,9\}$. The blocks\enlargethispage{1pc}
\begin{equation*}
\begin{array}{cccccc}
(6,7,6,9)& (3,4,6,1)& (9,3,7,5)& (9,1,0,9) & (5,2,9,6)& (7,1,7,3)\\
(1,2,1,8)& (5,1,5,4) & (8,0,1,6)& (4,7,0,4)& (8,4,3,9)& (8,5,7,8) \\
(2,0,7,2)& (9,4,8,2)& (6,2,4,5)& (6,3,0,8) & (0,5,0,3)& (3,3,2,3) \\
\end{array}
\end{equation*}
form a perfect $(4,2)_{10}\text{-DCC}$ of size~18.
\qquad\end{proof}

\begin{lemma}
\label{DL for 12} 
$DL(12,4)=24\in {\rm Spec}(12,4,2)$.
\end{lemma}

\begin{proof}
Take the alphabet $X$ to be $\bbZ_{12}$. The orbits of the eight codewords
\begin{equation*}
\begin{array}{cccc}
( 1, 2,10, 4)& (10, 3, 5,10)& ( 0,11,10, 8)& ( 4, 6,10, 9) \\
( 3, 6, 7, 3)& ( 5, 9, 7, 5)& (4, 1, 7, 8)& (4,11, 5, 4) \\
\end{array}
\end{equation*}
under the action of adding $4$ modulo $12$ form a perfect
$(4,2)_{12}\text{-DCC}$ of size~24.
\qquad\end{proof}

\begin{lemma}
\label{DL+1 for 12}
$DL(12,4)+1=25\in {\rm Spec}(12,4,2)$.
\end{lemma}

\begin{proof}
Take the alphabet $X$ to be $\bbZ_2\times\{0,1,2,3,4,5\}$.
The orbits of the 12 codewords
\begin{equation*}
\begin{array}{ccc}
((0,1),(0,1),(0,0),(1,1)) & ((0,2),(0,0),(0,1),(0,2)) & ((0,3),(1,0),(0,2),(0,3)) \\
((0,4),(0,1),(1,2),(0,4)) & ((0,5),(0,2),(1,2),(0,5)) & ((0,0),(0,3),(0,4),(0,5)) \\
((0,1),(0,3),(1,5),(1,3)) & ((0,2),(1,3),(1,1),(0,4)) & ((0,4),(0,0),(1,5),(1,4)) \\
((0,4),(1,3),(0,2),(1,0)) & ((0,5),(1,1),(1,5),(0,0)) & ((0,5),(1,4),(1,3),(0,1)) \\
\end{array}
\end{equation*}
under the action of $\bbZ_2\times\{0,1,2,3,4,5\}$, together with
the additional codeword $((0,0),$ $(1,0),(1,0),(0,0))$, form a
perfect $(4,2)_{12}\text{-DCC}$ of size~25.
\qquad\end{proof}

\begin{lemma}
\label{lem7.6}
$[26,30] \subseteq {\rm Spec}(12,4,2)$.
\end{lemma}

{\it Proof}.
There exists a $\{4\}$-DGDD of type $3^4$ by Lemma~\ref{GDDtoDGDD}
and Theorem~\ref{4GDDg^t}. Now apply Proposition~\ref{membership}
to obtain
\begin{align*}
{\rm Spec}(12,4,2) & \supseteq \{18\} + \sum_{i=1}^{4} {\rm Spec}(3,4,2) \\
& = \left[26,30\right].
\qquad\endproof
\end{align*}

\begin{lemma}
\label{lem7.7}
$\{30,37,38,39\}\subseteq {\rm Spec}(13,4,2)$.
\end{lemma}

\begin{proof}
Take a perfect ${\rm IDP}(13:1,4)$ (which exists by Lemma~\ref{13:1}) and fill in the
hole of size one with a perfect $(4,2)_1\text{-DCC}$ of size one to obtain a $(4,2)_{13}\text{-DCC}$
of size~30.

Now, take a ${\rm B}(13,4,1)$ (which exists by Theorem~\ref{Hanani}) and construct
a $\{4\}\text{-GDD}$ of type $1^9 4^1$ by considering a block as a group.
It follows from Lemma~\ref{GDDtoDGDD} that there exists a $\{4\}\text{-DGDD}$ of type $1^9 4^1$.
Applying Proposition~\ref{membership} then gives $37,38,39\in{\rm Spec}(13,4,2)$.
\qquad\end{proof}

\begin{lemma}
\label{lem7.8}
$DL(16,4)=44\in {\rm Spec}(16,4,2)$.
\end{lemma}

\begin{proof}
Take the alphabet $X$ to be $\bbZ_8\times\{0,1\}$. The orbits of the five codewords
\begin{equation*}
\begin{array}{ccc}
((0,0),(1,0),(1,1),(0,0)) & ((0,0),(4,1),(6,1),(7,1)) & ((0,1),(2,0),(7,1),(4,1)) \\
((0,1),(5,0),(0,1),(0,0)) & ((0,1),(6,1),(4,0),(0,0)) & \\
\end{array}
\end{equation*}
under the action of $\bbZ_8\times\{0,1\}$, together with the four additional codewords
\begin{equation*}
\begin{array}{cc}
((0,0),(2,0),(4,0),(6,0)) & ((6,0),(4,0),(2,0),(0,0)) \\
((1,0),(3,0),(5,0),(7,0)) & ((7,0),(5,0),(3,0),(1,0)) \\
\end{array}
\end{equation*}
form a perfect $(4,2)_{16}\text{-DCC}$ of size~44.
\qquad\end{proof}

\begin{lemma}
\label{lem7.9}
$[48,56] \subseteq {\rm Spec}(16,4,2)$.
\end{lemma}

{\it Proof}.
Apply Proposition~\ref{membership} to a $\{4\}\text{-DGDD}$ of
type $4^4$, which exists by Theorem~\ref{4DGDD}, to conclude that
\begin{align*}
{\rm Spec}(16,4,2) & \supseteq \{32\} + \sum_{i=1}^{4} {\rm Spec}(4,4,2) \\
& = \left[48,56\right].
\qquad\endproof
\end{align*}

\begin{lemma}
\label{lem7.10}
${\rm Spec}(28,4,2)=[133,154]=I(28,4)$.
\end{lemma}

\begin{proof}
Take the alphabet $X$ to be $\bbZ_{14}\times\{0,1\}$. The
orbits of the nine codewords
\begin{equation*}
\begin{array}{cc}
((0,0),(1,0),(0,0),(0,1)) & ((0,1),(2,0),(4,0),(0,1)) \\
((0,0),(3,0),(7,0),(12,0)) & ((0,1),(8,0),(12,1),(11,1)) \\
((0,1),(3,1),(5,1),(12,0)) & ((0,0),(2,1),(8,1),(8,0)) \\
((0,0),(5,1),(9,1),(6,0)) & ((0,0),(6,1),(1,1),(11,0)) \\
((0,0),(11,1),(7,1),(10,0)) \\
\end{array}
\end{equation*}
under the action of $\bbZ_{14}\times\{0,1\}$, together with the
seven additional codewords
\begin{equation*}
\begin{array}{cc}
((0,1),(7,1),(1,1),(8,1)) & ((2,1),(9,1),(3,1),(10,1)) \\
((4,1),(11,1),(5,1),(12,1)) & ((6,1),(13,1),(7,1),(0,1)) \\
((8,1),(1,1),(9,1),(2,1)) & ((10,1),(3,1),(11,1),(4,1)) \\
((12,1),(5,1),(13,1),(6,1)) & \\
\end{array}
\end{equation*}
form a perfect $(4,2)_{28}$-DCC of size~133.

To show that $[134,154]\subseteq {\rm Spec}(28,4,2)$,
apply Proposition~\ref{membership} to a $\{4\}\text{-DGDD}$ of type $7^4$, which exists
by Theorem~\ref{4DGDD}, to conclude that
\begin{align*}
{\rm Spec}(28,4,2) & \supseteq \{98\} + \sum_{i=1}^{4} {\rm Spec}(7,4,2) \\
& = [134,154],
\end{align*}
since ${\rm Spec}(7,4,2)=[9,14]$ and $\sum_{i=1}^4
\{9,10,11,12,13,14\}$ = $[36,56]$.
\qquad\end{proof}

\subsection{\boldmath The case $q\equiv 0\pmod{3}$}\
\label{sec7.3}

\begin{lemma}
\label{lem7.11}
$DL(q,4) \in {\rm Spec}(q,4,2)$ for all $q\equiv 0\pmod{6}$, $q\geq 24$.
\end{lemma}

\begin{proof}
For $q\geq 24$, apply Proposition~\ref{membership} to a
$\{4\}$-DGDD of type $6^{q/6}$, which exists by Theorem~\ref{4DGDD}.
\qquad\end{proof}

\begin{lemma}
\label{lem7.12}
$[DL(q,4)+1,DU(q,4)-(\left\lfloor \frac{q}{12}\right\rfloor+1)]
\subseteq {\rm Spec}(q,4,2)$ for all $q\equiv 0\pmod{6}$, $q\geq
30$.
\end{lemma}

{\it Proof}.
By Theorem~\ref{4GDD6^um^1}, there exist both a $\{4\}$-GDD of
type $6^{(q-6)/6} 3^1$ and a $\{4\}$-GDD of type $6^{(q-6)/6} 6^1
3^1$. Apply Lemma~\ref{average} to obtain a $\{4\}$-DGDD of type
$6^{(q-6)/6} 3^2$. Now apply Proposition~\ref{membership} to
obtain
\begin{align*}
{\rm Spec}(q,4,2) & \supseteq \left\{\frac{q(q-6)}{6}\right\}+\sum_{i=1}^{2}{\rm Spec}(3,4,2)+
 \sum_{i=1}^{(q-6)/6} {\rm Spec}(6,4,2)\\
& =\left[\frac{q^2}{6}+1,\frac{q^2+4q-6}{6}\right]\\
& =\left[DL(q,4)+1,DU(q,4)-\left(\left\lfloor \frac{q}{12}\right\rfloor+1\right)\right].
\qquad\endproof
\end{align*}

\begin{lemma}
\label{lem7.13}
$\left[DL(q,4),DU(q,4)-\left\lfloor \frac{q+3}{12}\right\rfloor\right]
\subseteq {\rm Spec}(q,4,2)$ for all $q\equiv 3$ {\rm (mod~6)},
$q\geq 27$.
\end{lemma}

{\it Proof}.
For $q\geq 27$, there exists a $\{4\}$-DGDD of type $6^{(q-3)/6}3^1$ by
Lemma~\ref{GDDtoDGDD} and Theorem~\ref{4GDD6^um^1}.
Now apply Proposition~\ref{membership} to obtain
\begin{align*}
{\rm Spec}(q,4,2) & \supseteq \left\{\frac{(q-3)^2}{6} \right\} + {\rm Spec}(3,4,2)+
\sum_{i=1}^{(q-3)/6} {\rm Spec}(6,4,2) \\
& = \left[\frac{q^2+3}{6},\frac{q^2+4q-3}{6}\right] \\
& =\left[DL(q,4),DU(q,4)-\left\lfloor \frac{q+3}{12}\right\rfloor\right].
\qquad\endproof
\end{align*}

\begin{lemma}
\label{lem7.14}
There exists a perfect {\rm DP}$(12t,4)$ of size belonging to
$[DU(12t,4)-6t,DU(12t,4)]$ for all $t\geq 1$.
\end{lemma}

\begin{proof} 
An ${\rm RDP}(12t,4)$ is known
\cite{Assafetal:2001}. The leave graph $\Gamma$ of this
${\rm RDP}(12t,4)$ is 1-regular. For each $j \in [0,6t]$, adding
 $j$ blocks $(a,a,b,b)$ for $j$ arcs $(a,b)\in E(\Gamma)$ and $12t-2j$ blocks
 $(c,c,c,d)$ and $(d,d,d,d)$ for the remaining arcs
$(c,d)\in E(\Gamma)$ to this ${\rm RDP}(12t,4)$ gives a perfect
${\rm DP}(12t,4)$ having $DU(12t,4)-j$ blocks.
\qquad\end{proof}

\begin{lemma}
\label{lem7.15}
There exists a perfect {\rm DP}$(12t+3,4)$ of size belonging to
$[DU(12t+3,4)-6t,DU(12t+3,4)]$ for all $t\geq 1$.
\end{lemma}

{\it Proof}.
Let $(\{0,1,\ldots,12t+1\},\G,\A)$ be a $\{4\}$-GDD of type
$2^{6t+1}$, which exists by Theorem~\ref{4GDDg^t}, with
$\G=\{\{2i,2i+1\}:0\leq i\leq 6t\}$, and let
$(\{0,1,\ldots,12t+3\},\B)$ be a B$(12t+4,4,1)$, which exists by
Theorem~\ref{Hanani}, such that $[12t,12t+3]\in\B$. Let
$\C=\B\setminus\{[12t,12t+3]\}$, and let $\C(x)=\{A\in\C:x\in A\}$.
Then $\C(12t+2)$ and $\C(12t+3)$ are disjoint. Define
\begin{align*}
\A_1 & = \{A_\uparrow: A\in\A \}, \\
\A_2 & = \{A_\downarrow: \text{$A\in\C\setminus(\C(12t+2)\cup\C(12t+3))$}\}, \\
\A_3 & = \{(12t+2,a,b,c):\text{$a>b>c$ and $\{a,b,c,12t+2\}\in\C(12t+2)$}\}, \\
\A_4 & = \{(a,b,c,12t+2):\text{$a>b>c$ and $\{a,b,c,12t+3\}\in\C(12t+3)$}\}.
\end{align*}
Then $(\{0,1,\ldots,12t+2\},\cup_{i=1}^4 \A_i)$ is an
RDP$(12t+3,4)$ of size
$|\A|+(|\B|-1)=2t(6t+1)+(4t+1)(3t+1)-1=3t(8t+3)=RDU(12t+3,4)$.

For each $j \in [0,6t]$, to obtain a perfect DP$(12t+3,4)$ of
size $DU(12t+3,4)-j$ with this RDP$(12t+3,4)$ as a subsystem, we
add to it the following $12t+3-j$ blocks:
\begin{align*}
& (12t,12t,12t+1,12t+2), \\
& (12t+2,12t+1,12t+1,12t), \\
& (12t+2,12t+2,12t+2,12t+2), \\
& (2i,2i,2i+1,2i+1) \text{ for $0\leq i\leq j-1$, } \\
& (2i,2i,2i,2i+1) \text{ for $j\leq i\leq 6t-1$, and} \\
& (2i+1,2i+1,2i+1,2i+1) \text{ for $j\leq i\leq 6t-1$.}
\qquad\endproof
\end{align*}

\begin{lemma}
\label{7.16}
There exists a perfect {\rm DP}$(12t+6,4)$ of size belonging to
$[DU(12t+6,4)-6t,DU(12t+6,4)]$ for all $t\geq 1$.
\end{lemma}

\begin{proof}
Let $(\{0,1,\ldots,12t+7\},\G,\A)$ be a $\{4\}$-GDD of type
$2^{6t+4}$, which exists by Theorem~\ref{4GDDg^t}. Without loss of
generality, assume that $[12t+4,12t+7]\in\A$, and let
$(\{0,1,\ldots,12t+3\},\B)$ be a B$(12t+4,4,1)$, which exists by
Theorem~\ref{Hanani}. Let $\C=\A\setminus\{[12t+4,12t+7]\}$, and
let $\C(x)=\{A\in\C: x\in A\}$. Then $\C(12t+4)$, $\C(12t+5)$,
$\C(12t+6)$, and $\C(12t+7)$ are pairwise disjoint. Define
\begin{align*}
\A_1 & = \{A_\uparrow:\text{$A\cap[12t+4,12t+7]=\text{\O}$ and $A\in\C$}\}, \\
\A_2 & = \{(12t+4,a,b,c):\text{$a<b<c$ and $\{a,b,c,12t+4\}\in\C(12t+4)$}\}, \\
\A_3 & = \{(12t+5,a,b,c):\text{$a<b<c$ and $\{a,b,c,12t+5\}\in\C(12t+5)$}\}, \\
\A_4 & = \{(a,b,c,12t+5):\text{$a<b<c$ and $\{a,b,c,12t+6\}\in\C(12t+6)$}\}, \\
\A_5 & = \{(a,b,c,12t+4):\text{$a<b<c$ and $\{a,b,c,12t+7\}\in\C(12t+7)$}\}, \\
\A_6 & = \{B_\downarrow:\text{$B\in\B$}\}.
\end{align*}
Then $(\{0,1,\ldots,12t+5\},\cup_{i=1}^6\A_i)$ is an
RDP$(12t+6,4)$ of size
$(|\A|-1)+|\B|=2(3t+2)(2t+1)-1+(4t+1)(3t+1)=24t^2+21t+4=RDU(12t+6,4)$.

For each $j \in [0,6t]$, to derive a perfect DP$(12t+6,4)$ of
size $DU(12t+6,4)-j$ with this RDP$(12t+6,4)$ as a subsystem, we
add $12t+6-j$ blocks to this RDP$(12t+6,4)$ as follows. Partition
the set $\{\{a,b\}\in\G, a<b<12t+4\}$ into two disjoint parts
$\G_0$ and $\G_1$ such that $|\G_0|=j$. Hence, $|\G_1|=6t-j$. Let
\begin{align*}
\B_0 & = \{(a,a,b,b):\text{$\{a,b\}\in\G_0$}\}, \\
\B_1 & = \{(a,a,a,a),(a,b,b,b):\text{$\{a,b\}\in\G_1$}\}, \\
\B_2 & = \{(a,a,a,12t+4): \text{$\{a,12t+4\}\in\G$}\}, \\
\B_3 & = \{(a,a,a,12t+5): \text{$\{a,12t+5\}\in\G$}\}, \\
\B_4 & = \{(a,a,a,12t+5): \text{$\{a,12t+6\}\in\G$}\}, \\
\B_5 & = \{(a,a,a,12t+4): \text{$\{a,12t+7\}\in\G$}\}, \text{ and} \\
\B_6 & = \{(a,a,a,b),(b,b,b,a):
\text{$\{a,b\}=\{12t+4,12t+5\}$}\}.
\end{align*}
Then $(\{0,1,\ldots,12t+5\},\B_0\cup_{i=1}^6(\A_i\cup\B_i))$ is a
perfect DP$(12t+6,4)$ of size $DU(12t+6,4)-j$.
\qquad\end{proof}

\begin{lemma}
\label{lem7.17}
There exists a perfect {\rm DP}$(12t+9,4)$ of size belonging to
$[DU(12t+9,4)-(6t+2),DU(12t+9,4)]$ for all $t\geq 1$.
\end{lemma}

\begin{proof}
Let $(\{0,1,\ldots,12t+7\},\G,\A)$ be a $\{4\}$-GDD of type
$2^{6t+4}$ with $\{12t+2i,12t+2i+1\}\in\G$ for $i\in[3]$, which
exists by Theorem~\ref{4GDDg^t}, and let
$(\{0,1,\ldots,12t+9\},\B)$ be a PBD$(12t+10,\{4,7^\star\})$ with
$[12t+3,12t+9]$ as the block of size seven, which exists by
Theorem~\ref{47*}. Let $\C=\B\setminus\{[12t+3,12t+9]\}$, and let
$\C(x)=\{A\in\C: x\in A\}$. Define
\begin{align*}
\A_1 & = \{A_\uparrow:\text{$A\in\A$}\}, \\
\A_2 & = \{(12t+7,12t+6,12t+8,12t+5), (12t+8,12t+6,12t+7,12t+4), \\
& \ \phantom{= \{}(12t+5,12t+4,12t+3,12t+8)\}, \\
\A_3 & = \{A_\downarrow:\text{$A\in\C\setminus\C(12t+9)$}\}, \\
\A_4 & = \{(a,b,c,12t+8):\text{$a>b>c$ and $\{a,b,c,12t+9\}\in\C(12t+9)$}\}.
\end{align*}
Then $(\{0,1,\ldots,12t+8\},\cup_{i=1}^4\A_i)$ is an
RDP$(12t+9,4)$ of size
$|\A|+(|\B|+2)=2(3t+2)(2t+1)+(4t+1)(3t+4)+3=RDU(12t+9,4)$.

For each $j \in [0,6t+2]$, to derive a perfect DP$(12t+9,4)$ of
size $DU(12t+9,4)-j$ with this RDP$(12t+9,4)$ as a subsystem, we
add $12t+9-j$ blocks to this RDP$(12t+9,4)$ as follows. Partition
the set $\{\{a,b\}\in\G, a<b<12t+4\}$ into two disjoint parts
$\G_0$ and $\G_1$ such that $|\G_0|=j$. Hence, $|\G_1|=6t+2-j$.
Let
\begin{align*}
\B_0 & = \{(a,a,b,b):\text{$\{a,b\}\in\G_0$}\}, \\
\B_1 & = \{(a,a,a,b), (b,b,b,b):\text{$\{a,b\}\in\G_1$}\}, \\
\B_2 & = \{(12t+i,12t+i,12t+i,12t+3): i\in\{6,7,8\}\}, \\
\B_3 & = \{(12t+4,12t+4,12t+4,12t+5)\}, \text{ and} \\
\B_4 & = \{(12t+5,12t+5,12t+5,12t+4)\}.
\end{align*}
Then $(\{0,1,\ldots,12t+8\},\B_0\cup_{i=1}^4(\A_i\cup\B_i))$ is a
perfect DP$(12t+9,4)$ of size $DU(12t+9,4)-j$.
\qquad\end{proof}

We summarize these results as follows.

\begin{theorem}
\label{thm7.18}
${\rm Spec}(q,4,2)=I(q,4)$ for all $q\equiv 0$ {\rm (mod~3)},
except when $q=6$, and except possibly when $q\in\{9, 15, 18,
21, 24\}$.
\end{theorem}

\subsection{\boldmath The case $q\equiv 1\pmod{3}$}\
\label{sec7.4}

\begin{lemma}
\label{lem7.19}
${\rm Spec}(q,4,2)=I(q,4)$ for all $q\equiv 1$ {\rm (mod~12)},
$q\geq 49$.
\end{lemma}

\begin{proof} 
Take a ${\rm B}(13,4,1)$ (which exists by Theorem~\ref{Hanani})
and construct a $\{4\}\text{-GDD}$ of type $1^9 4^1$ by
considering a block as a group. It follows from Lemma~\ref{GDDtoDGDD} that there exists a $\{4\}\text{-DGDD}$ of type
$1^9 4^1$. Applying Proposition~\ref{membership} then gives an
incomplete code with $x$ codewords,
where $x\in\{36,37,38\}$, each with a hole
of size one. We also have a perfect ${\rm IDP}(13:1,4)$ (which
exists by Lemma~\ref{13:1}).

We first deal with $q\ge 61$. By Theorem~\ref{4GDD12^um^1}, there
exist both a $\{4\}$-GDD of type $12^{{q-13 \over 12}}6^1$ and a
$\{4\}$-GDD of type $12^{{q-13 \over 12}} 12^1 6^1$. Apply Lemma~\ref{average} to obtain a $\{4\}$-DGDD of type $12^{{q-13 \over
12}} 6^2$. Add one point and apply Proposition~\ref{membership-1}.
For every group of size~12, we put a copy of an incomplete code on
13 points with $x$ codewords,
where $x\in\{29, 36, 37, 38\}$, each having a hole
of size one. For one of the group of size six, we put a copy of an
incomplete code on seven points with $x$ codewords,
where $x\in\{9,10,11,12,13\}$,
each having a hole of size one. For the last group of size six, we
put a $(4,2)_{7}$-DCC of size $x$, where $x \in [9,14]$. This
gives the entire spectrum from $DL(q,4)$ to $DU(q,4)$.

For $q=49$, take a $\{4\}$-DGDD of type $12^{4}$ and add one
point, fill in the groups of size~12 with a $(4,2)_{13}$-DCC
having a hole of size one and 29 codewords. Add one codeword for
the hole of size one. This gives a $(4,2)_{49}$-DCC with $DL(49,4)$
codewords. Now, take a $\{4\}$-DGDD of type $7^7$ (which exists by
Theorem~\ref{4DGDD}). Fill in the groups with a $(4,2)_{7}$-DCC
having $9,10,11,12,13,$ or $14$ codewords; this gives the
remaining cases.
\qquad\end{proof}

\begin{lemma}
\label{lem7.20}
${\rm Spec}(q,4,2)=I(q,4)$ for all $q\equiv 4$ {\rm (mod~12)},
$q\geq 64$.
\end{lemma}

\begin{proof} 
For the sizes belonging to $[DL(q,4)+1, DL(q,4)+3]$, take a
$\{4\}$-DGDD of type $12^{{q-4 \over 12}}3^1$ (which exists by
Theorem~\ref{4GDD12^um^1} and Lemma~\ref{GDDtoDGDD}) and add one
point. For every group of size~12, we put a copy of a perfect
${\rm IDP}(13:1,4)$ with 29 blocks. For the last group, we put a
$(4,2)_{4}$-DCC of size~$x$, where $4 \leq x \leq 6$.

For the sizes belonging to $[DL(q,4)+4, DU(q,4)]\cup\{DL(q,4)\}$,
take a $\{4\}$-DGDD of type $12^{{q-16 \over 12}}15^1$ (which
exists by Theorem~\ref{4GDD12^um^1} and Lemma~\ref{GDDtoDGDD}) and
add one point. For every group of size~12, we put a copy of an
incomplete code on 13 points with $x$ codewords,
where $x\in\{29,36,37,38\}$,
each having a hole of size one. For the last group, we put a
$(4,2)_{16}$-DCC of size~$x$, where $x \in
\{44,48,49,50,51,52,53,54,$\break$55,56\}$\mbox{.\qquad}\end{proof}

\begin{lemma}
\label{lem7.21}
${\rm Spec}(q,4,2)=I(q,4)$ for all $q\equiv 7$ {\rm (mod~12)},
$q\geq 55$.
\end{lemma}

\begin{proof} 
For the sizes belonging to $[DL(q,4)+1, DL(q,4)+5]$, take a
$\{4\}$-DGDD of type $12^{{q-7 \over 12}}6^1$ (which exists by
Theorem~\ref{4GDD12^um^1} and Lemma~\ref{GDDtoDGDD}) and add one
point. For every group of size~12, we put a copy of a perfect
${\rm IDP}(13:1,4)$ with 29 blocks. For the last group, we put a
$(4,2)_{7}$-DCC of size~$x$, where $9 \leq x \leq 14$.

Take a $\{4\}$-DGDD of type $6^5$ (which exists by Theorem~\ref{4DGDD}), 
add one point, and apply Proposition~\ref{membership-1}. For four groups of size~six, we put a copy of an
incomplete code on seven points with $x$ codewords,
where $x\in\{9,10,11,12,13\}$,
each having a hole of size one. For the last group, we put a
$(4,2)_{7}$-DCC of size~$x$, where $x \in \{9,10,11,12,13,14\}$.
This gives $[DL(31,4)+2, DU(31,4)]=[165,186]\subseteq {\rm
Spec}(31,4,2)$.

When $q \ge 103$, for the sizes belonging to $[DL(q,4)+6,
DU(q,4)]$, take a $\{4\}$-DGDD of type $12^{{q-31 \over 12}}30^1$
(which exists by Theorem~\ref{4GDD12^um^1} and Lemma~\ref{GDDtoDGDD}) and add one point. For every group of size~12, we
put a copy of an incomplete code on 13~points with $x$ codewords,
where $x\in\{29,36,37,38\}$, each having a hole of size one. For the last
group, we put a $(4,2)_{31}$-DCC of size~$x$, where $x \in
[165,186]$.

When $q\in\{55, 67,79,91\}$, for the sizes belonging to
$[DL(q,4)+6, DU(q,4)]$, we take a $\{4\}$-DGDD of type $6^9,
6^{11}, 6^{13},$ or $6^{13} 12^1$ (which exists by Theorem~\ref{4GDD6^um^1} and Lemma~\ref{GDDtoDGDD}), respectively, to
obtain the desired codes. Here, we fill in one group of size six
with a code on seven points and other groups with incomplete
codes.
\qquad\end{proof}

\begin{lemma}
\label{lem7.22}
${\rm Spec}(q,4,2)=I(q,4)$ for all $q\equiv 10$ {\rm (mod~12)}, $q\geq 58$.
\end{lemma}

\begin{proof} 
Take a $\{4\}$-DGDD of type $12^{{q-10 \over 12}}9^1$ (which exists by
Theorem~\ref{4GDD12^um^1} and Lemma~\ref{GDDtoDGDD}) and add one
point. For every group of size~12, we put a copy of a perfect
${\rm IDP}(13:1,4)$ with 29 blocks. For the last group, we put a
$(4,2)_{10}$-DCC of size~$18$. This gives a $(4,2)_{q}$-DCC with
$DL(q,4)$ codewords.

By Theorem~\ref{4GDD12^um^1}, there exist both a $\{4\}$-GDD of
type $12^{{q-10 \over 12}}3^1$ and a $\{4\}$-GDD of type
$12^{{q-10 \over 12}} 12^1 3^1$. Apply Lemma~\ref{average} to
obtain a $\{4\}$-DGDD of type $12^{{q-10 \over 12}} 6^1 3^1$. Add
one point and fill in the holes. For every group of size~12, we
put a copy of an incomplete code on 13 points with $x$ codewords,
where $x\in\{29,36,37,38\}$, each having a hole of size one. For the group of
size six, we put a copy of an incomplete code on seven points with $x$ codewords,
where $x\in\{9,10,11,12,13\}$, each having a hole of size one.
For the last group of size three, we put a $(4,2)_{4}$-DCC of size
$x$, where $x \in \{4,5,6\}$. This gives the entire spectrum from
$DL(q,4)+1$ to $DU(q,4)$.
\qquad\end{proof}

We summarize these results as follows.

\begin{theorem}
\label{thm7.23}
${\rm Spec}(q,4,2)=I(q,4)$ for all $q\equiv 1$ {\rm (mod~3)},
except when $q=4$, and except possibly when $q\in\{10, 13, 16,
19, 22, 25, 31, 34, 37, 40, 43, 46, 52\}$.
\end{theorem}

\subsection{\boldmath The case $q\equiv 2\pmod{3}$}\
\label{sec7.5}

\begin{lemma}
\label{lem7.24}
${\rm Spec}(q,4,2)=I(q,4)$ for all $q\equiv 2\pmod{6}$.
\end{lemma}

{\it Proof}.
The case $q=2$ is settled by Lemma~\ref{smallspec4lemma}.
For $q\geq 8$, apply Proposition~\ref{membership} to a $\{4\}$-DGDD of type $2^{q/2}$, which exists by
Theorem~\ref{4DGDD}, to conclude that
\begin{align*}
{\rm Spec}(q,4,2) & \supseteq \left\{ \frac{q(q-2)}{6}\right\} + \sum_{i=1}^{q/2} {\rm Spec}(2,4,2) \\
& = \left[\frac{q(q+1)}{6},\frac{q(q+4)}{6}\right] \\
& = I(q,4).
\qquad\endproof
\end{align*}

To deal with $q\equiv 5\pmod{6}$, we need a class of $\{4\}$-DGDDs
that we establish below.

\begin{lemma}
\label{4DGDD2^t5^1} 
A $\{4\}$-{\rm DGDD} of type $2^{t}5^1$ exists
if and only if $t\equiv 0\pmod{3}$, $t\geq 6$.
\end{lemma}

\begin{proof}
We first establish necessity of the condition $t\equiv 0\pmod{3}$,
$t\geq 6$. Suppose $x$ is a point contained in a group of size
two. Then $x$ must appear with each of $2t+3$ points twice. Hence
$2(2t+3)\equiv 0\pmod{3}$. This gives $t\equiv 0\pmod{3}$. Any
$\{4\}$-DGDD must contain at least four groups, so a $\{4\}$-DGDD
of type $5^1$ cannot exist. The nonexistence of a $\{4\}$-DGDD of
type $2^35^1$ is easily established by computation. Hence $t\geq 6$.

To construct a $\{4\}$-DGDD of type $2^6 5^1$, take a resolvable
$(\{3\},2)$-GDD $(X,\G,\A)$ of type $2^6$, which exists by Theorem~\ref{3RGDD}, and let $\{\A_1,\ldots,\A_{10}\}$ be a partition of
$\A$ into parallel classes. This GDD is directable by Theorem~\ref{directable} and hence underlies a $\{3\}$-DGDD $(X,\G,\B)$ of
type $2^6$. Now let $\infty_i\not\in X$, $i\in[5]$, and form $\C$
as follows:
\begin{align*}
\C & = \{ (a,b,c,\infty_i): \text{$(a,b,c)\in\B$ and $\{a,b,c\}\in \A_i$}\} \\
 & \cup 
 \{ (\infty_i,a,b,c):\text{$(a,b,c)\in\B$ and $\{a,b,c\}\in \A_{i+5}$}\}.
\end{align*}
Then $(X,\G\cup\{\{\infty_1,\ldots,\infty_5\}\},\C)$ is a
$\{4\}$-DGDD of type $2^65^1$.

For $t\geq 9$, a $\{4\}$-DGDD of type $2^t5^1$ exists by Lemma~\ref{GDDtoDGDD} and Theorem~\ref{4GDD2^t5^1}\mbox{.\qquad}\end{proof}

\begin{lemma}
\label{lem7.26}
${\rm Spec}(q,4,2)=I(q,4)$ for all $q\equiv 5\pmod{6}$, $q\not=11$.
\end{lemma}

{\it Proof}.
The case $q=5$ is settled by Lemma~\ref{smallspec4lemma}. For
$q\geq 17$, apply Proposition~\ref{membership} to a $\{4\}$-DGDD
of type $2^{(q-5)/2}5^1$, which exists by Lemma~\ref{4DGDD2^t5^1},
to conclude that
\begin{align*}
{\rm Spec}(q,4,2) & \supseteq \left\{\frac{(q+3)(q-5)}{6}\right\} + \sum_{i=1}^{(q-5)/2} {\rm Spec}(2,4,2) + {\rm Spec}(5,4,2) \\
& = \left[\frac{q(q+1)}{6}, \frac{q^2+4q-3}{6}\right]\\
& = I(q,4).
\qquad\endproof
\end{align*}

We summarize these results as follows.

\begin{theorem}
\label{thm7.27}
${\rm Spec}(q,4,2)=I(q,4)$ for all $q\equiv 2\pmod{3}$, except possibly when $q=11$.
\end{theorem}

\subsection{Summary}\
\label{sec7.6}

\begin{theorem}
\label{thm7.28}
${\rm Spec}(q,4,2)=I(q,4)$ for all positive integers $q$, except
when $q \in \{4, 6\}$, and except possibly when $q\in\{9$, $10$, $11$,
$13$, $15$, $16$, $18$, $19$, $21$, $22$, $24$, $25$, $31$, $34$, $37$, $40$, $43$, $46$,
$52\}$.
\end{theorem}

\section{Conclusion}
\label{sec8}

Much work has been done on the existence of a perfect
$(n,n-2)_q\text{-DCC}$. However, very little is known on the
general problem of determining the spectrum of possible sizes for a
perfect $(n,n-2)_q\text{-DCC}$. In this paper, we determine
completely the spectrum of possible sizes for perfect $q$-ary
1-deletion-correcting codes of length three for all $q$, and
perfect $q$-ary 2-deletion-correcting codes of length four for
all but $19$ values of $q$. A complete solution to these
undetermined codes appears difficult.

\end{document}